\documentclass{aa}
\usepackage{graphicx}

\usepackage{txfonts}

\usepackage{natbib}
\bibpunct{(}{)}{;}{a}{}{,}

\begin{document}

\title{Star formation activity of intermediate redshift cluster galaxies out to the infall regions}

\author{B. Gerken\inst{1}\fnmsep\thanks{Visiting astronomer of the
               German--Spanish Astronomical Center, Calar Alto, operated
               by the Max--Planck--Institut f\"ur Astronomie, Heidelberg,
               jointly with the Spanish National Commission for Astronomy.}
               \fnmsep\thanks{\emph{Present address:}University of Oxford,
                        Astrophysics, Denys Wilkinson Building, Keble Road, 
                        Oxford OX1 3RH, UK}
 \and B. Ziegler\inst{1}\fnmsep$^{\star}$
 \and M. Balogh\inst{2}  
 \and D. Gilbank\inst{3} 
 \and A. Fritz\inst{1}\fnmsep$^{\star}$
 \and K. J\"ager\inst{1}}

\titlerunning{Star formation activity out to the cluster infall regions}
\authorrunning{B. Gerken et al.}

\offprints{B. Gerken, \email{bettinag@astro.ox.ac.uk}}

\institute{Universit\"atssternwarte G\"ottingen, 
           Geismarlandstr. 11, 
           37083 G\"ottingen, Germany 
 \and
           Department of Physics, University of Durham, 
           South Road, 
           Durham DH1 3LE, UK
 \and 
           Department of Astronomy and Astrophysics, University of Toronto,
	   60 St. George Street,
	   Toronto, Ontario, Canada M5S 3H8}
\date{Received / Accepted }

\abstract{
We present a spectroscopic analysis of two galaxy clusters at z$\approx0.2$, out to $\sim4$\,Mpc. The two clusters VMF73 and VMF74 as identified by \citet{VMFJQH98} were observed with multiple object spectroscopy using MOSCA at the Calar Alto 3.5\,m telescope. Both clusters lie in the \mbox{ROSAT} Position Sensitive Proportional Counter field R285 and were selected from the X-ray Dark Cluster Survey \citep{GBCZ04} that provides optical $V$- and $I$-band data. VMF73 and VMF74 are located at respective redshifts of z$=0.25$ and z$=0.18$ with velocity dispersions of 671 km\,s$^{-1}$ and 442 km\,s$^{-1}$, respectively. Both cluster velocity dispersions are consistent with Gaussians. The spectroscopic observations reach out to $\sim2.5$ virial radii. Line strength measurements of the emission lines H$_\alpha$ and [O\,II]$\lambda$3727 are used to assess the star formation activity of cluster galaxies which show radial and density dependences. The mean and median of both line strength distributions as well as the fraction of star forming galaxies increase with increasing clustercentric distance and decreasing local galaxy density. Except for two galaxies with strong H$_\alpha$ and [O\,II] emission, all of the cluster galaxies are normal star forming or passive galaxies. Our results are consistent with other studies that show the truncation in star formation occurs far from the cluster centre.

}

\maketitle

\section{Introduction} \label{sec:Intro}
Galaxy properties such as colour, morphology and spectral characteristics are functions of redshift as well as of galaxy environment. Examples for cluster specific redshift dependences are the Butcher-Oemler effect \citep{BO78a} and the change of the ``morphological mix'' in clusters with redshift. Local clusters like the Coma cluster are dominated by S0 galaxies, which form only a small fraction in higher redshift clusters at z$\sim0.5$, whereas spiral galaxies are more abundant in distant than in local clusters. This observation has raised the question whether the spiral galaxies in distant clusters are the progenitors of S0 galaxies in local clusters, and which kind of processes may be responsible for a transformation of one morphological type into another \citep[e.g.][]{DOCSE97,PSDCB99}. Further evidence for an evolution of galaxy properties with redshift is provided by studies which show that the universal average star formation rate (SFR) has been rapidly declining since z$\sim$1 \citep[e.g.][]{LLHD96,MFDGS96,BSIK99,SPF01}. However, at a given epoch, cluster galaxies always show suppressed star formation compared with the field population at the same redshift \citep[e.g.][]{BMYCE97}. Models of hierarchical structure formation predict a continuous accumulation of material and substructure to more and more massive galaxies, groups and clusters as time progresses \citep[e.g.][ and references therein]{Kauff96,KCDW99b,CLBF00}. Therefore, it may be possible to link the decline in the global star formation rate to the growth of large scale structure in the universe.\\
A number of studies have investigated the connection between the cluster environment and morphological gradients \citep[e.g.][]{Dress80,PG84,WG93,DOCSE97,DML01}. To characterize the cluster environment, the clustercentric distance of a galaxy as well as the local projected galaxy density are useful as independent parameters. Both of them, however, are projected parameters. The morphological dependence of cluster galaxies on the environment has been studied based on the clustercentric radius e.g. by \citet{WG91,WG93} and based on local galaxy density by \citet{Dress80,DOCSE97} and others. It has been discussed which parameter is more appropriate, i.e. whether galaxy morphology is influenced rather by global properties, characterized by the clustercentric distance, or by local properties like the local galaxy density. However, in centrally concentrated, regular and presumably relaxed clusters the local galaxy density is closely correlated with the clustercentric radius \citep{DOCSE97,GNMBG03}. A correlation between morphological gradients and clustercentric distance or local galaxy density has been found, such that the dense cluster core is populated predominantly by ellipticals and S0 galaxies. Towards the cluster outskirts, the fraction of spirals increases while the elliptical and S0 fraction decreases \citep{Dress80}.\\
In a similar way, the star formation activity of galaxies depends on the environment. This dependence has been investigated by various studies, which reveal a lower level of star formation in cluster galaxies compared with the field population at the same redshift \citep{ASHCY96,BSMYC98,BBSZD02}. The results of these studies suggest that star formation is suppressed in cluster galaxies over a wide range of cluster masses. To a certain extent it is not surprising to find lower SFRs of galaxies in high density environments as these dense regions like cluster cores are dominated by galaxy types with an intrinsically lower SFR. However, recent studies show that even within the same galaxy type the SFR is lower in the cluster population compared to the surrounding field \citep{BSMYC98,GNMBG03}.\\
The present analysis focuses on star formation properties of galaxies in clusters at z\,$\sim0.2$ out to large distances from the cluster centre. Studies of \mbox{CNOC\,1} data explored clusters out to $\sim1$ virial radius (R$_v$) and found that mechanisms responsible for the low SFR of cluster galaxies are likely to start acting far out of the cluster core in the infall region, where field galaxies attracted by the cluster potential experience the influence of the cluster environment for the first time \citep{BMYCE97,BSMYC98}. It has been shown that galaxies as far out as $\sim2$\,R$_{v}$ may have crossed the cluster core within a Hubble time \citep{RS98,BNM00}. This suggests that galaxies in the cluster outskirts may have undergone significant evolution on their way through the cluster core and that cluster specific processes start acting on infalling galaxies even beyond 2\,R$_v$. It has been discussed, for instance, whether the cluster environment induces starbursts in infalling galaxies which lead to a rapid consumption of the gas supply and a subsequent passive evolution of these galaxies \citep{DG83,BD86,ASHCY96}. \\
In the local universe, the cluster environment has been explored from the densest central regions to large radial distances and into the field by studies of \mbox{2dFGRS} and \mbox{SDSS} data \citep{LBDCB02,GNMBG03,BEMLB04}. At intermediate and high redshift, only few comparable studies have yet been conducted \citep{ASHCY96,KSNOB01}. Galaxy properties, however, depend strongly on redshift. Therefore, to understand the evolution of galaxies, it is essential to investigate different environments at different redshifts. In this study we present observations of two low-mass clusters at z$\sim0.2$ as part of a larger program to study the correlation between star formation and environment at this redshift.\\
The outline of the paper is as follows. In chapter 2, the selection of objects, observations and data reduction are described. The data analysis and results are presented in chapter 3. In chapter 4, the results are discussed and compared with previous studies, and chapter 5 is a summary of our findings. Throughout this paper, we use a cosmology of H$_0=70$\,km\,s$^{-1}$\,Mpc$^{-1}$, $\Omega_m=0.3$ and $\Omega_{\Lambda}=0.7$

\section{Data}
\subsection{Observations}
\subsubsection{Photometry}
The photometric data were obtained during two observing runs in June 1998 and January 1999 with the Wide Field Camera at the Isaac Newton Telescope, La Palma. These observations formed the optical basis of the X-ray Dark Cluster Survey \mbox{(XDCS)}, an optical follow up survey of X-ray imaging for galaxy clusters \citep{GBCZ04}. The aim of this survey was a comparison between cluster finding algorithms based on the optical and X-ray luminosities of clusters. For this purpose, deep archival \mbox{ROSAT} Position Sensitive Proportional Counter \mbox{(PSPC)} fields part of which had been analysed by \citet{VMFJQH98} were imaged in the optical $V$- and $I$-band.

\subsubsection{Object Selection for Spectroscopy} \label{sec:ObjSel}
Three $40\arcmin \times40 \arcmin$ fields, each of which contains two galaxy clusters from the \mbox{XDCS} catalogue \citep{GBCZ04}, were selected for spectroscopic observations. In the present study, the two clusters contained in the \mbox{ROSAT} \mbox{PSPC} field R285 are analysed. The J2000-coordinates of the X-ray centroids of these clusters and their estimated redshifts and X-ray fluxes \citep{VMFJQH98} are given in Table~\ref{tab:clcoor}. In order to enhance the observational efficiency, fields with two clusters aligned along the line of sight were chosen. The selection of the spectroscopic targets was directed by three criteria. Firstly, galaxies were selected based on the $I$-magnitude only, i.e. no colour criterion was applied. This restriction avoids a bias in selecting against certain galaxy types, as there is a correlation of galaxy type with colour. Secondly, in the selection procedure priority was given to the brighter objects. The third criterion was a spatial constraint to the mask. The slitlets had to have a parallel orientation with respect to each other and to the mask orientation. Furthermore, each slit needs a minimum length to provide a sufficient sky background subtraction. All selection effects resulting from these criteria were accounted for with the selection function (see Sec.~\ref{sec:Selfct}).  

\subsubsection{Spectroscopic Observations} \label{sec:Spec}
\begin{figure*}
\centering
\includegraphics[width=1.0\textwidth]{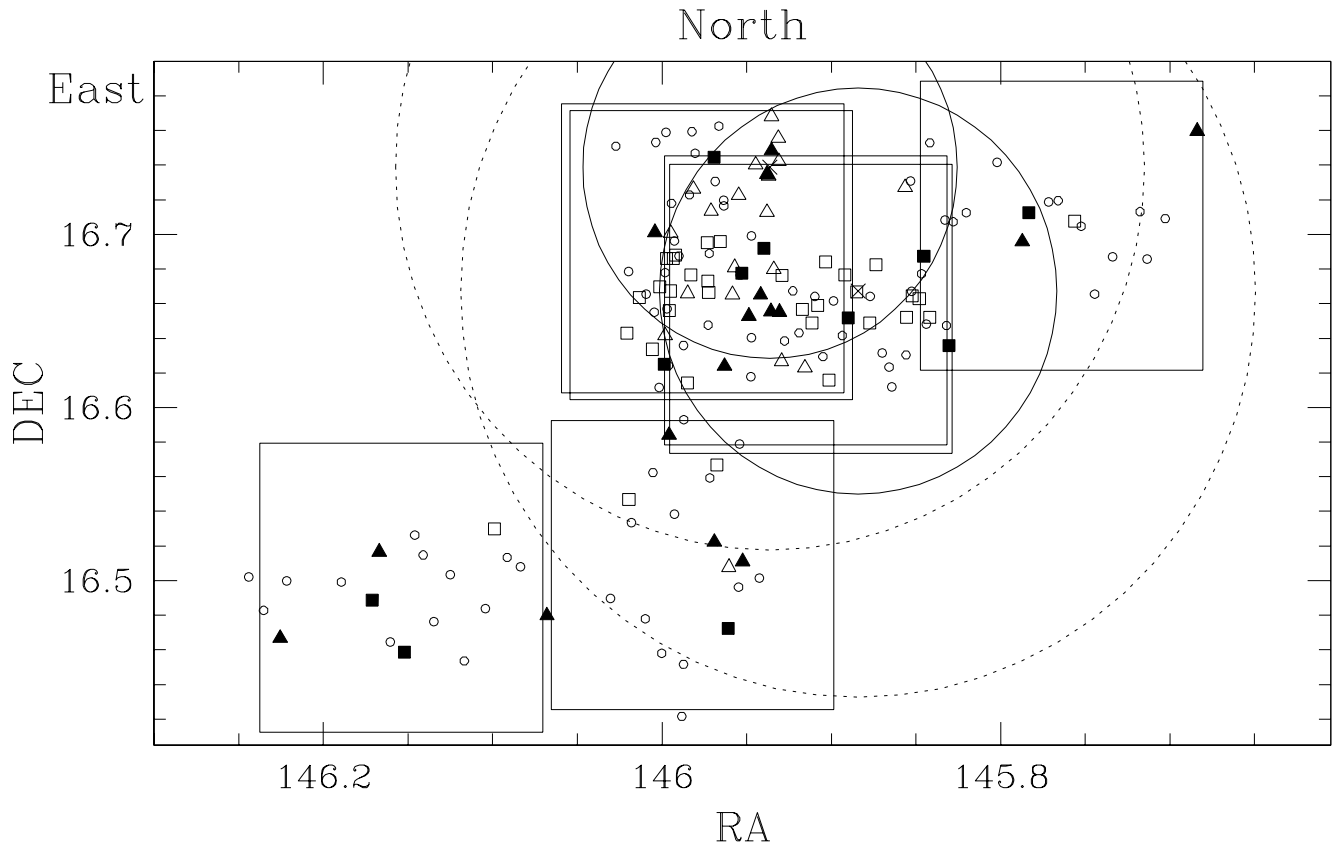}
\caption[Cluster Positions]{\small Positions of the two clusters in field R285. Squares and triangles represent member galaxies of VMF73 and VMF74, respectively, empty circles are spectroscopically observed objects which do not belong to any cluster. Open squares and triangles represent all cluster galaxies, while filled symbols show only members with W$_0$([O\,II])\,$>5$\,\AA\ (line strength measurements are described in section~\ref{sec:WoMeas}). The large crosses show the respective cluster X-ray centroids. The solid and dotted circles show one and two virial radii, respectively, of VMF73 and VMF74. The seven slitmasks covering the two galaxy clusters are shown as large squares.}
\label{fig:Posmas}
\end{figure*}

The spectroscopic observations were carried out on 2002 February 10--15 at the 3.5\,m telescope on the Calar Alto observatory. During these 5 nights and additional service mode time each field was observed with 7 to 8 slitmasks, each mask containing 20 to 30 slits for simultaneous spectroscopy. Each slit had a length of about 10 to 20 arcseconds, to ensure an accurate sky background subtraction. For the multiobject spectroscopic (MOS) observations, the focal reducer MOSCA was used. This instrument is installed at the Ritchey--Chr\`etien focus of the 3.5\,m telescope and allows multiobject spectroscopy placing a multiple slit mask as aperture into the focal plane. For our observations the grism \texttt{green\_500} was used which has a central wavelength of 5500\,\AA\ and a typical wavelength range of 4300\,\AA\ to 8200\,\AA\@, thus being ideally suited for observations of the spectral lines H$_{\alpha}$ and [O\,II]\,$\lambda$3727 at a redshift of $z\approx0.2$. Grism \texttt{green\_500} has a mean dispersion of 2.9\,\AA\@/pixel. The exposure times for each mask ranged between one hour and three hours, depending on the apparent magnitude of the objects. Their total $I$-band magnitude ranged between 16.7 and 19.5. The total observing time amounted to $\sim48$ hours. In total, 553 spectra were taken, including 8 objects which turned out to be stars. The two clusters in field R285 analysed in the present paper were observed with seven slitmasks as shown in figure~\ref{fig:Posmas}. 
The dense central parts were observed with four masks which cover the respective cluster area within $\sim1$\,R$_v$. The outer regions contain galaxies beyond $\sim1$\,R$_v$ and were observed with three masks.
\begin{table}
\centering
\begin{tabular}{c c c c c}
Coordinates & Cluster & z$_{VMF}$ & f$_X$\,[$10^{14}$\,ergs\,s$^{-1}$\,cm$^{-2}$]\\
\hline
  9:43:32 16:40:02 & VMF73  & 0.256 & 23.1\\
  9:43:45 16:44 20 & VMF74  & 0.180 & 21.2\\
\end{tabular}
\caption[Coordinates of observed fields]{\small J2000--coordinates, estimated redshifts and X--ray flux as given by \citet{VMFJQH98} of \mbox{ROSAT} \mbox{PSPC} field R285} 
\label{tab:clcoor}
\end{table}

\subsection{Data Reduction}
The data reduction was carried out with the image processing system \texttt{MIDAS}\footnote{MIDAS is provided by the European Southern Observatory (ESO).} as described in \citet{ZBSDL01} and \citet{FZBSD04}. Reduction steps included bias subtraction, correction of the spatial distortion, flatfielding, sky subtraction, optimal extraction of one-dimensional spectra, wavelength calibration and redshift determination. The correction for the spatial distortion (rectification) was achieved by fitting a third order polynomial to the spectrum. Each line of the spectrum was reproduced ten times in order to enable subpixel shifts. Subsequently the spectrum was rectified row by row and rebinned to the original size. Due to the strong spatial distortion of the spectra, the flatfielding had to be done after the rectification. In almost all cases the individual slit images had to be extracted including some light contamination from neighbouring slits in order not to loose spectral information in the spectrum. This light contamination could be removed only after the rectification, therefore the flatfielding had to be be carried out afterwards as well. The sky background was subtracted by fitting a polynomial of first order to each column separately, in some cases a fixed value was subtracted. The wavelength calibration was done with HgArNe calibration frames and the rms of the solution ranged between 0.1 and 0.3 \AA\ at a central wavelength of 5500\,\AA\@. Galaxy redshifts were determined by centering prominent emission or absorption lines with Gaussians and calculating the mean value. Deviations in the determinations from line to line ranged between 100 and 250 km/s which is a reasonable accuracy for the spectral resolution. The spectral lines were identified visually which was a problem in cases of low signal to noise ratio. For 39 spectra of low signal, a reliable redshift determination was not possible; these spectra were excluded from further analysis. For 8 further spectra the redshift determination was ambiguous. In that case the spectrum was reexamined with the \texttt{IRAF}\footnote{IRAF is distributed by NOAO which is operated by AURA Inc. under contract with NSF} routine \texttt{fxcor}. This routine is based on an algorithm of \citet{TD79} and performs a Fourier cross-correlation on the object and to one or several zero redshift template spectra. For 5 out of the 8 spectra with ambiguous z the redshift determination could be improved.

\subsection{Selection Function} \label{sec:Selfct}
\begin{figure}
\begin{minipage}{0.45\linewidth}
\centering
\includegraphics[width=1.0\textwidth]{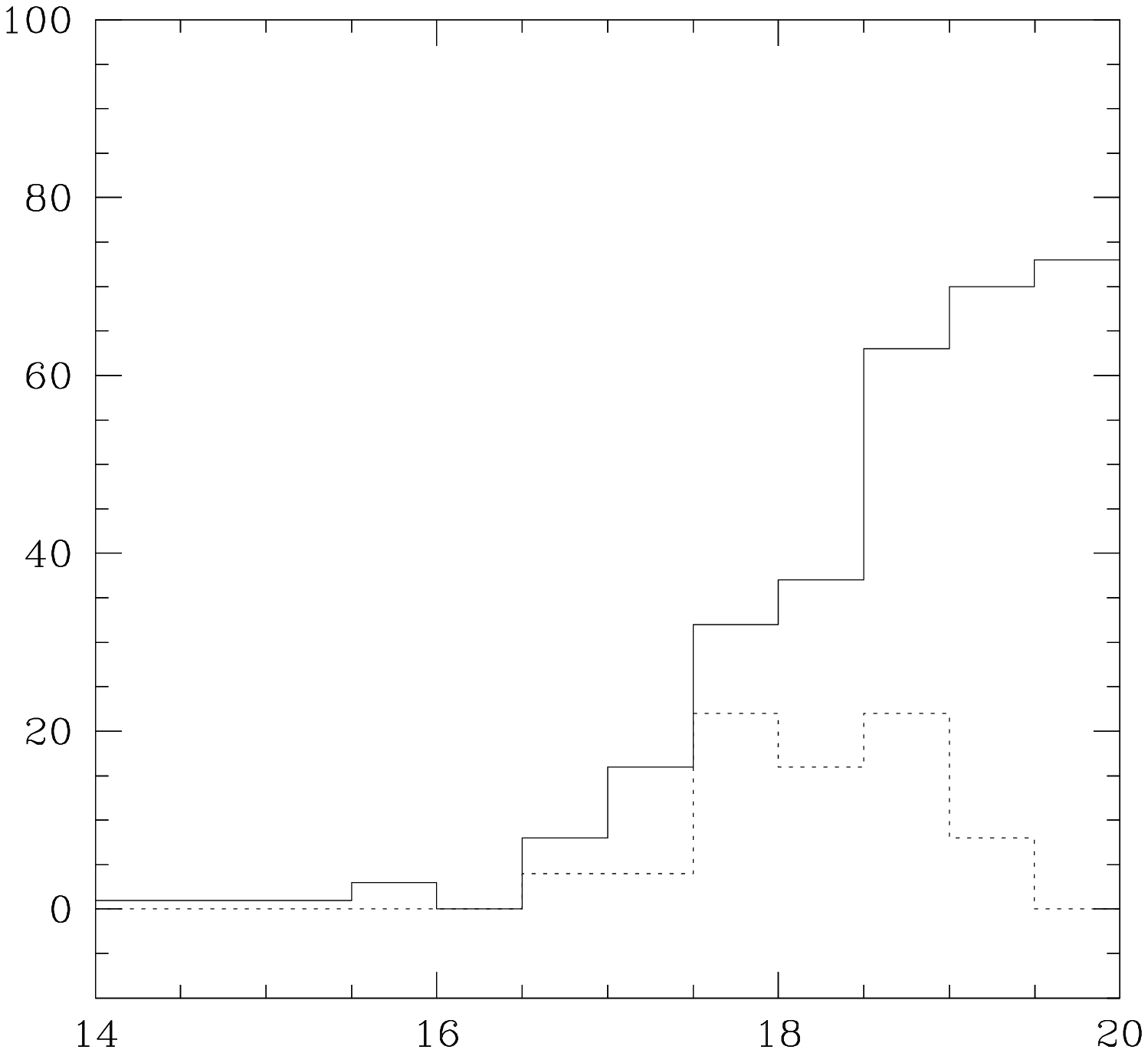}
\end{minipage}
%\hspace*{0.1cm}
\begin{minipage}{0.45\linewidth}
\centering
\includegraphics[width=1.0\textwidth]{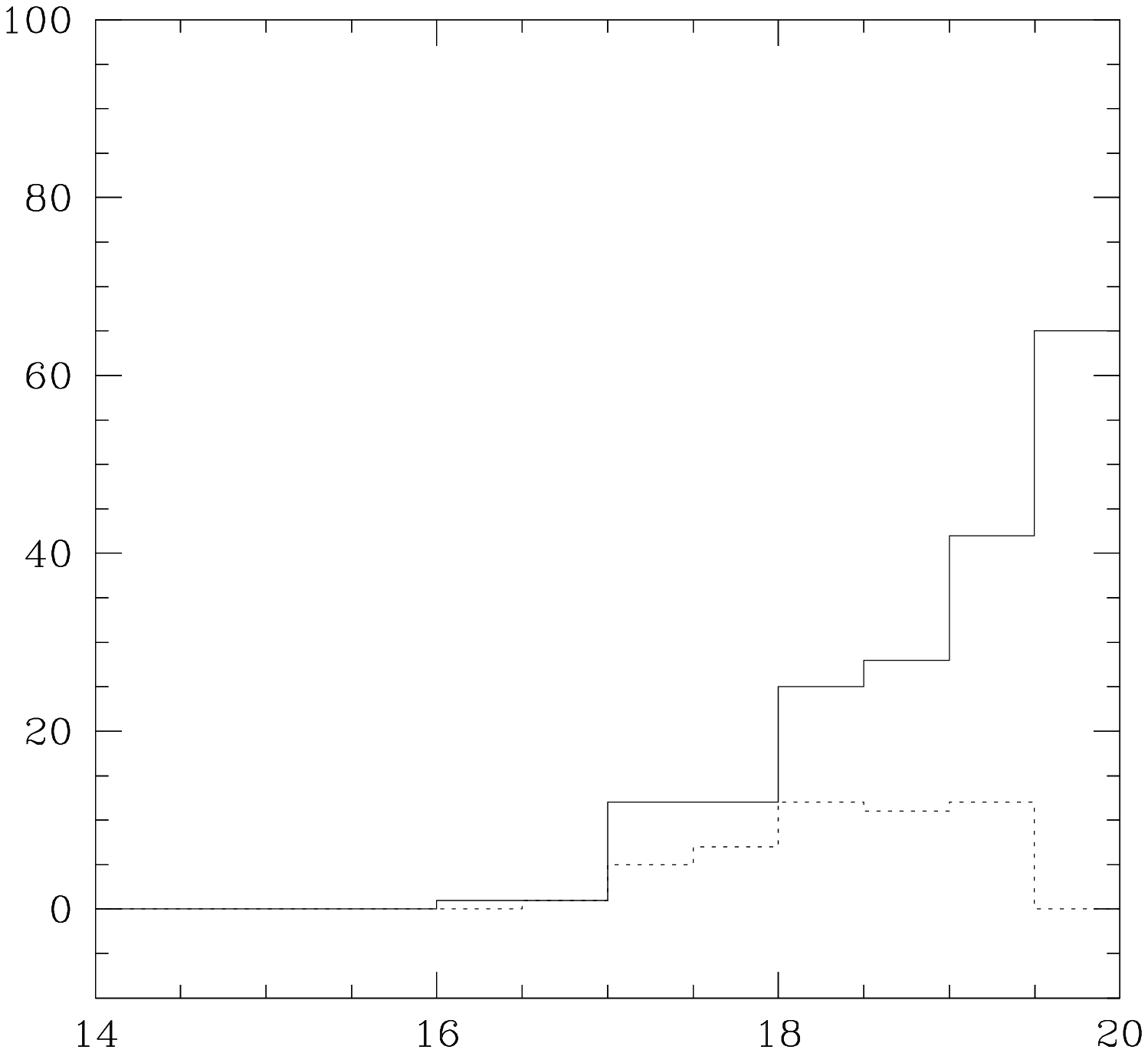}
\end{minipage}
\begin{minipage}{0.45\linewidth}
\centering
\includegraphics[width=1.0\textwidth]{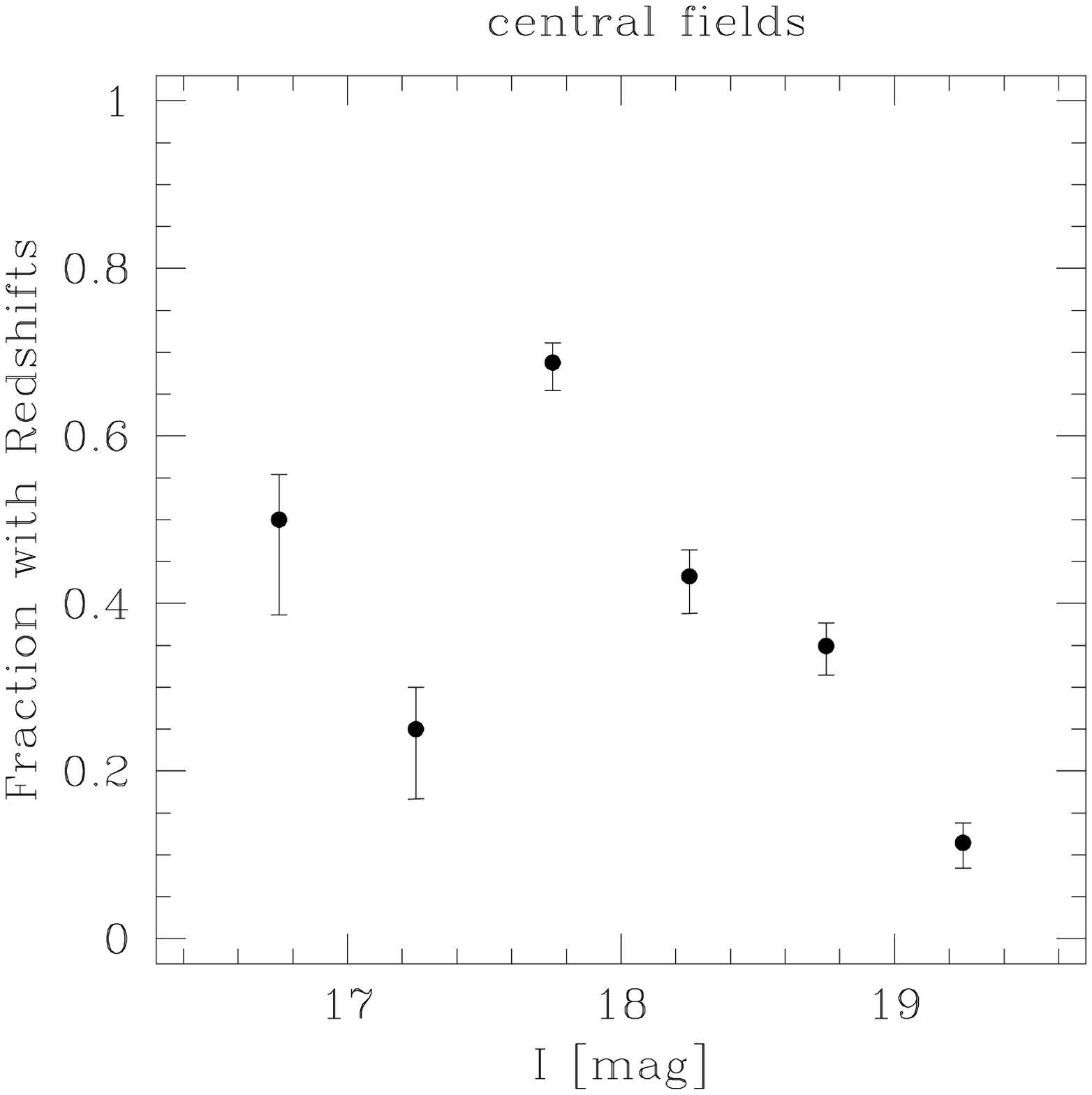}
\end{minipage}
\hspace*{0.4cm}
\begin{minipage}{0.45\linewidth}
\centering
\includegraphics[width=1.0\textwidth]{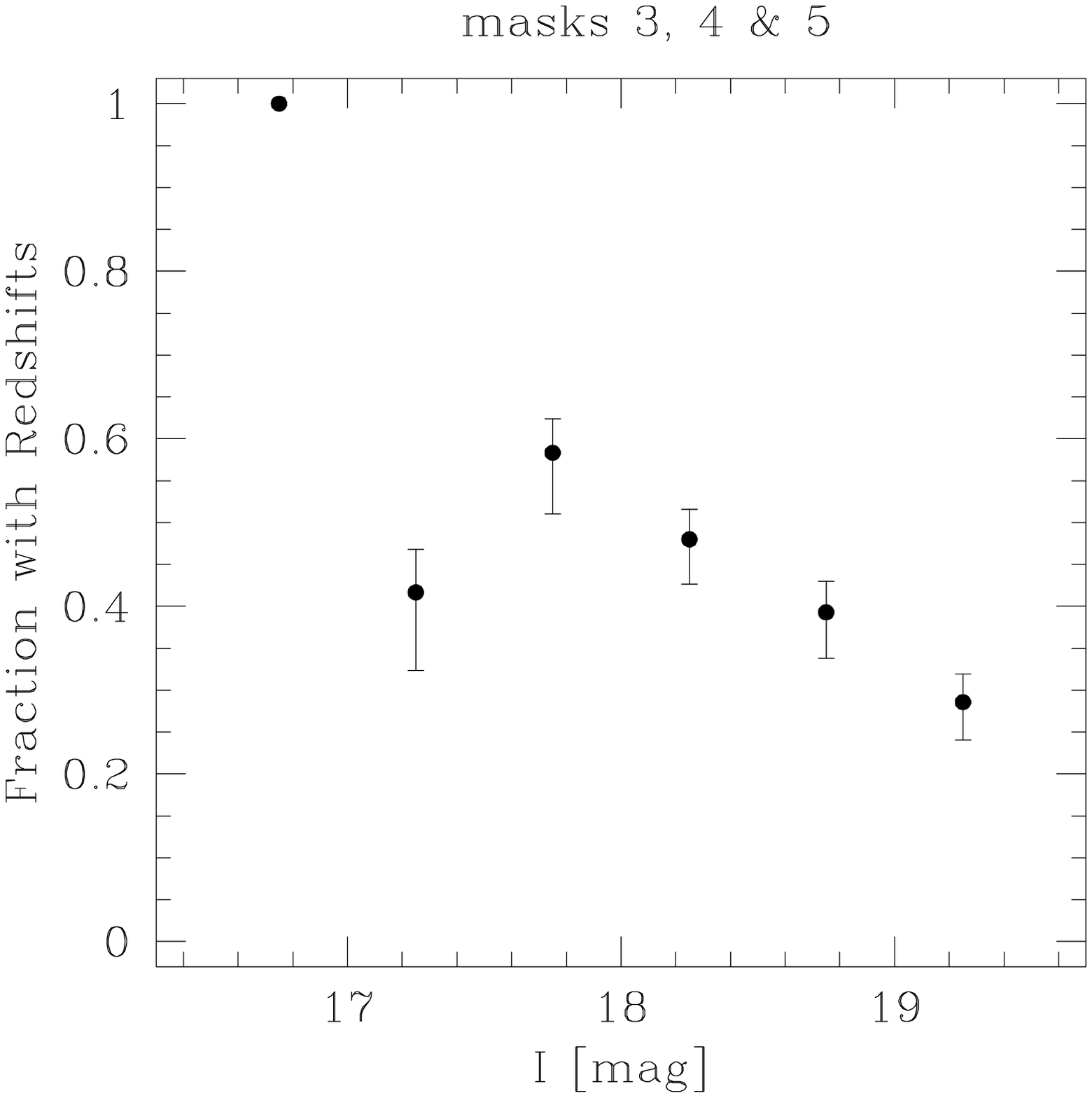}
\end{minipage}
\caption[Galaxy distribution in field R285 as a function of observed magnitude]{\small \textbf{Top left: }Galaxy distribution in the area covered by the four central masks in field R285 as a function of observed magnitude. The solid line represents all objects in the photometric catalogue, the dotted line shows the distribution of objects with confirmed redshift. The fraction of galaxies with redshift is calculated as the ratio of both distributions in each magnitude bin. The bins have a size of 0.5\,mag. The faint end of the dotted distribution corresponds to the magnitude limit of I$_{tot}\,=$\,19.5 of the spectroscopic sample. \textbf{Top right: }The same as in the top left panel, but for the area covered by the three outer masks. \textbf{Bottom left: }Selection function in the area covered by the central masks. Each point shows the fraction of galaxies with redshifts in the corresponding magnitude bin. Error bars are Poisson statistics. \textbf{Bottom right: }The same as in the bottom left panel, but for the outer masks.} 
\label{fig:selfct}
\end{figure}

In both clusters of field R285, only part of all member galaxies was observed spectroscopically. Therefore, selection effects have to be corrected for which is achieved by a selection function. The selection function is defined as the fraction of galaxies with a reliable redshift in the photometric catalogue and is a function of the apparent magnitude. The selection effect is not uniform throughout the cluster due to two effects. Firstly, the central regions of VMF73 and VMF74 were observed with four masks overlapping each other, while the outer regions were observed with three masks without overlap. This leads to a significantly higher coverage and thus to a higher fraction of objects with spectroscopy in the central regions. Secondly, in the central regions the galaxy concentration is higher which leads to a higher fraction of spectroscopically observed objects in the centre. In the outer regions with lower galaxy density there are less objects available for spectroscopy. Both effects were accounted for by establishing two separate selection functions; one for the central cluster fields and another one for the outer regions. This correction is independent of cluster membership, i.e. no background and foreground corrections were applied. All objects with spectroscopy were divided into an inner and an outer sample according to the slit masks they were observed with. The four masks covering roughly the area out to 1\,R$_v$ of the respective clusters are considered the central masks, while the three masks covering the area beyond $\sim1$\,R$_v$ are referred to as the outer masks (see Figure~\ref{fig:Posmas}). For each of the two zones the selection function was calculated as follows. In a first step, the distribution of all galaxies in the photometric catalogue as a function of apparent $I$-band magnitude was determined. This distribution is shown as solid line in the top of Figure~\ref{fig:selfct}. 
Each magnitude bin has a size of 0.5\,mag. As a next step, the distribution of galaxies with reliable redshifts in the respective cluster region was determined, shown as dotted line in the top of Figure~\ref{fig:selfct}. The fraction of galaxies with redshifts as a function of apparent $I$ magnitude, i.e. the selection function, was calculated as the ratio of both distributions in each magnitude bin. The faint end of the dotted distribution corresponds to the magnitude limit in the $I$ band of $I_{tot}\,=$\,19.5 of the spectroscopic sample. The selection functions for the two cluster regions are shown in the bottom of Figure~\ref{fig:selfct}.\\  
The selection function was obtained based on the observed $I$-band magnitude. However, in order to correct for selection effects, it has to be applied to the quantity in question like absolute magnitude, [O\,II] line strength etc. Fractions and absolute numbers of these quantities were therefore corrected by assigning each galaxy a weight number which was the inverse of the corresponding selection function. This correction was applied seperately in the inner and the outer regions; the division into the inner and outer samples was performed according to the slit masks.

\section{Analysis and Results}
\subsection{Cluster Mean Redshifts} \label{sec:MnRed}
\begin{figure}
\centering
\resizebox{\hsize}{!}{\includegraphics{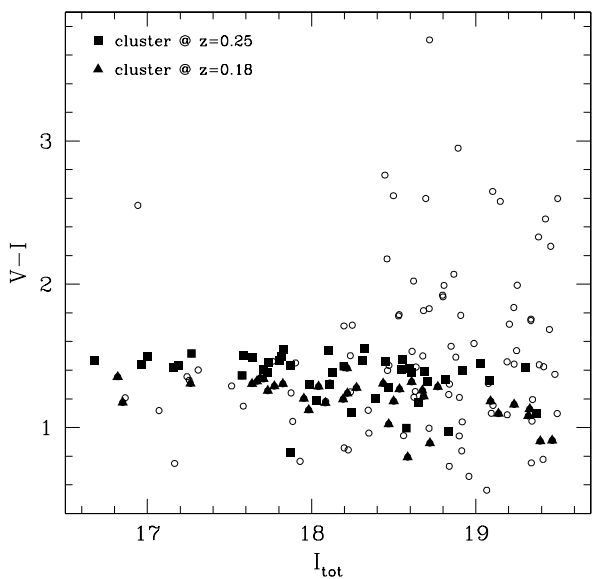}}
\caption[Colour--Magnitude--Diagram of the two clusters VMF73 and VMF74]{\small Colour--Magnitude--Diagram for the two clusters VMF73 and VMF74 in the field R285. Triangles and squares represent cluster members of VMF74 and VMF73, respectively Open circles are objects with spectroscopy which do not belong to either cluster.}
\label{fig:cmd}
\end{figure}

The Colour--Magnitude--Diagram (CMD) for the objects with spectroscopy is plotted in Figure~\ref{fig:cmd}. 
Triangles and squares represent galaxies of the two clusters VMF74 and VMF73, respectively. Open circles show objects with spectroscopy which do not belong to either cluster. Two distinct red sequences of early-type galaxies are distinguishible, corresponding to the two clusters along the line of sight. The large number of spectroscopically observed objects which are redder than the cluster galaxies on the red sequences and which do not belong to either cluster is due to the selection criteria. As outlined in Sec.~\ref{sec:ObjSel}, the selection was not based on colours in order to avoid biases against galaxy colours and types.\\
\begin{figure}
\resizebox{\hsize}{!}{\includegraphics{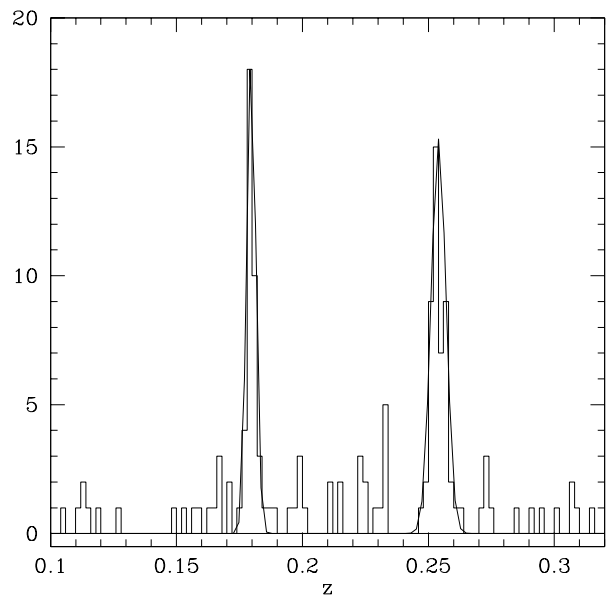}}
\caption[Redshift distributions of clusters VMF74 and VMF73]{\small Redshift distribution of field R285. Two peaks are visible indicating two clusters at redshifts of z$\approx$0.18 and z$\approx$0.25, respectively. The overlayed curves are Gaussians centered on the cluster mean redshift with a $\pm1$\,$\sigma$ velocity dispersion.}
\label{fig:histz}
\end{figure}
The redshift distribution of galaxies in the field R285 is shown in Figure~\ref{fig:histz}. Galaxies without a reliable redshift determination and stars have not been considered for this plot. Two peaks appear in the distribution, centered at redshifts of z$\approx0.18$ and z$\approx0.25$. This distribution clearly indicates the presence of two clusters in field R285, projected onto each other on the sky sphere. Both distributions are consistent with Gaussians, which are overlayed on the cluster mean redshift with a $\pm1$\,$\sigma$ velocity dispersion. All galaxies within 3\,$\sigma$ of the respective cluster mean redshift are considered as cluster members. The cluster mean redshifts were calculated with the biweight estimator as described in \citet{BFG90}. The calculation was performed for each cluster separately with an initial redshift range of $0.15\leq$\,z\,$\leq0.21$ for VMF74 and $0.22\leq$\,z\,$\leq0.28$ for VMF73. Both mean redshifts were computed in three iterations. The velocity dispersion of both clusters was calculated with the biweight estimator of scale. Like the biweight location estimator, the scale estimator was computed in three iterations to determine the velocity dispersion of VMF73 and VMF74. The mean redshifts, velocity dispersions, the number of spectroscopic members within 3\,$\sigma$, bolometric X-ray luminosities and estimated virial radii (see Sec.~\ref{sec:Dyn}) of both clusters are summarized in Table~\ref{clinfo}. For our cosmology, the distance moduli and the scale factor for the conversion from arcseconds to kpc are 40.54 and 3.96 for VMF73, respectively and 39.70 and 3.03 for VMF74, respectively.
\begin{table}
\centering
\begin{tabular}{c c c c c c c c}
\hline
Name & z & $\sigma$\,[km\,s$^{-1}$] & N & L$_X$\,[ergs\,s$^{-1}$] & R$_v$ [kpc]\\
\hline 
VMF73 & 0.254 & 671 & 44 & 6.03$\times10^{43}$ & 1672 \\
VMF74 & 0.180 & 442 & 35 & 2.35$\times10^{43}$ & 1207 \\
\hline
\end{tabular}
\caption[Redshifts of clusters VMF73 and VMF74]{Cluster IDs, Redshifts, velocity dispersions, number of members within 3\,$\sigma$, bolometric X-ray luminosity and virial radius of the two clusters contained in field R285.} 
\label{clinfo}
\end{table}

\subsection{Dynamics} \label{sec:Dyn}
\begin{figure}
\resizebox{\hsize}{!}{\includegraphics{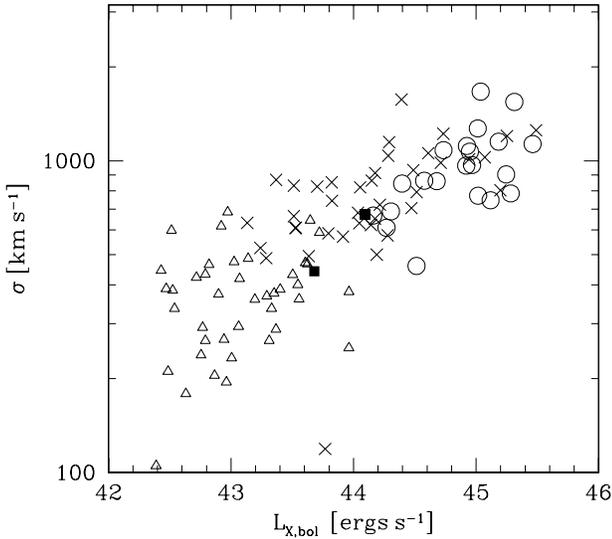}}
\caption[Correlation between bolometric L$_X$ and velocity dispersion]{\small Correlation between the bolometric X-ray luminosity and the velocity dispersion $\sigma$. Open triangles, crosses and circles represent local samples of \citet{XW00,DSJFV93} and \citet{Marke98}, respectively. The two clusters VMF73 and VMF74 are shown as solid squares.}
\label{fig:Lxsig}
\end{figure}

The measured velocity dispersions are in good agreement with the relation between the bolometric X-ray luminosities and velocity dispersions of local cluster samples, as shown in Figure~\ref{fig:Lxsig}.  
The open triangles, circles and crosses represent local samples of \citet{XW00,Marke98} and \citet{DSJFV93}, respectively, while the clusters VMF73 and VMF74 are represented by filled squares. The L$_X$-$\sigma$ correlation holds over a wide range of X-ray luminosities and velocity dispersions, reflecting a mass range from rich clusters to groups. Based on the relation between mass and velocity dispersion, the virial radius can be inferred \citep{GGMMB98}: 
\begin{equation} \label{eq:rvir}
R_{v}[Mpc]\sim 0.0035(1+z)^{-1.5}\sigma[km\,s^{-1}]
\end{equation}
Thus, the virial radii are $\sim1.21$\,Mpc or $\sim6'\,38''$ for VMF74 and $\sim1.67$\,Mpc or $\sim7'\,02''$ for VMF73. Henceforward, galaxies of both clusters are combined to one sample in order to increase the statistics. \\
The respective $I$ band brightest galaxies in both clusters have a certain offset from the X-ray centroid as given by \citet{VMFJQH98}. In VMF73, this offset is $6'\,36''$ from the X-ray centroid, while in VMF74 the offset is $1'\,29''$. Furthermore, in VMF74 the $I$ band brightest galaxy and the $V$ band brightest galaxy of the spectroscopic sample are two different objects. As the coverage of the brightest objects in the central fields is roughly 50\% (see Sec.~\ref{sec:Selfct}), it is possible that the two $I$ band brightest objects in the spectroscopic sample are not the respective brightest cluster galaxies of VMF73 and VMF74. Due to this uncertainty and as the X-ray luminosity is a better tracer of the gravitational potential than galaxies, the cluster centre is assumed to be the X-ray centroid as given by \citet{VMFJQH98} if not otherwise stated. The clustercentric radius of each cluster galaxy was calculated as the projected distance from the respective cluster X-ray centroid and normalised to the virial radius. 

\subsection{Radial and Density Dependences of Galaxy Properties}
\subsubsection{Measurements of Emission Line Strengths} \label{sec:WoMeas}
Star formation activity is closely linked with spectral properties such as line strengths of the Balmer series and forbidden oxygen lines. For this analysis, the emission lines H$_{\alpha}$ and the forbidden line [O\,II]$\lambda$3727 were employed as primary tracers of ongoing star formation. Both lines are indicative of hot, young O and B stars which ionize the surrounding gas in a star forming H\,II region. \\
The spectra of VMF73 and VMF74 member galaxies were not flux calibrated. We measure the equivalent widths of H$_{\alpha}$ and [O\,II] which are insensitive to this calibration. For the line strength measurements we adopt the line feature definitions of \citet{BMYCE99}. This system employs a central bandpass of the feature and two sidebands in which the continuum is measured. The midpoints in both sidebands are connected by a straight line and the flux is summed above the so defined continuum.

\subsubsection{Passive, Star Forming and Starburst Galaxies}
The level of star formation activity in a galaxy can be assessed by the emission line strengths of H$_{\alpha}$ and [O\,II]. Both lines were used for the analysis. In some member galaxies of cluster VMF74 with the lowest radial velocities the H$_{\alpha}$ line was affected by the atmospheric A-band, in such a way that part of the blue pseudocontinuum of H$_{\alpha}$ overlapped with the A-Band. This led to systematically higher values of the H$_{\alpha}$ line strength in the affected spectra and was corrected for in all cases by linearly interpolating the A-band. Passive, star forming and starburst galaxies show a wide range of line strengths in H$_\alpha$, [O\,II] and other emission and absorption lines \citep{Kenni92b}. In order to quantify the star formation activity of galaxies in the clusters VMF73 and VMF74, these three groups of galaxies will be discussed in the following. The classification of galaxy star formation properties is based on the MORPHS characterization \citep{PSDCB99}. The lower limit of 5\,\AA\ in the line strengths was chosen because any line strength detection below 5\,\AA\ cannot be distinguished from noise. This limit is adopted throughout the entire analysis. In order to allow a comparison with other studies at intermediate or higher redshift, for the following discussion of star formation properties mainly the [O\,II] line strength will be considered. \\
\textbf{Passive Galaxies}\\
Galaxies with W$_0$([O\,II])\,$<5$\,\AA\ are considered as passive, i.e. with no current star formation. Negative values of [O\,II] are caused by noise and features in the pseudocontinua. Within 3\,$\sigma$, both clusters VMF73 and VMF74 together have 79 member galaxies. A fraction of 69\%, weighted by the selection function, are classified as passive in the full sample. However, this fraction changes strongly with the distance from the centre: While in the inner regions the weighted fraction of passive galaxies is 80\%, it decreases to 23\% in the outer regions.\\
\textbf{Star Forming Galaxies}\\
Galaxies with ongoing star formation at the epoch of observation are all those with 5\,\AA\,$\leq$\,W$_0$([O\,II])\,$\leq$\,50\,\AA. The upper limit of 50\,\AA\ was adopted from \citet{Kenni92b} to account for a wide range of star formation rates in local spiral galaxies. In the two clusters VMF73 and VMF74, a total of 31\% of the member galaxies, weighted by the selection function, are found to have current star formation. Dividing the sample into an inner and an outer subsample, the weighted fraction of star forming galaxies in the inner regions is 20\% and 77\% in the outer regions.\\
\textbf{Starburst Galaxies}\\
This type of galaxy is undergoing a phase of intense star formation at the time of observation, which is characterized by an [O\,II] line strength of W$_0$([O\,II])\,$>50$\,\AA\@. In the cluster data, two candidates for starburst galaxies were detected.\\

\subsubsection{Emission Line Strengths as a Function of clustercentric Radius} \label {sec:W0R_VMF}
\begin{figure}
\resizebox{\hsize}{!}{\includegraphics{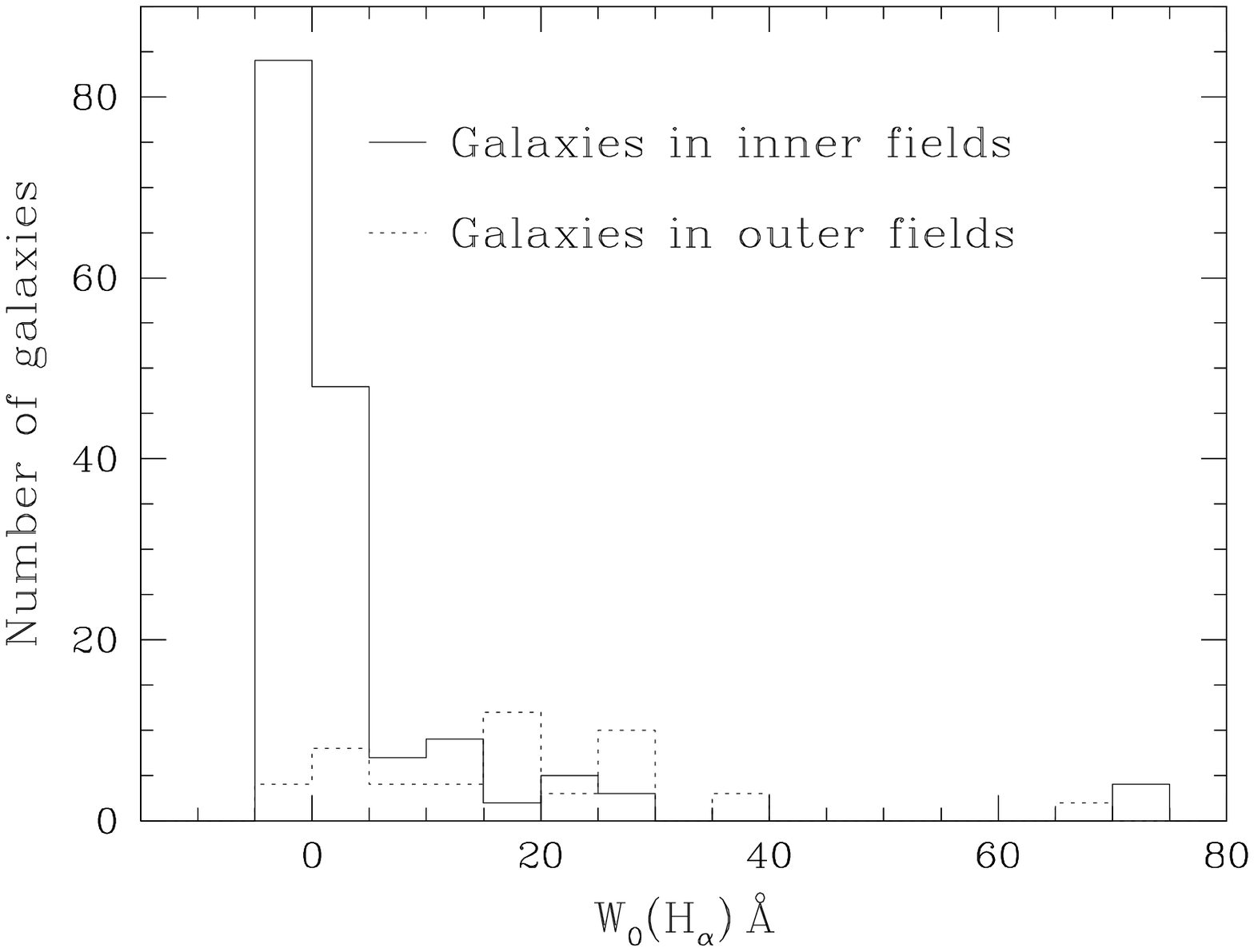}}
\resizebox{\hsize}{!}{\includegraphics{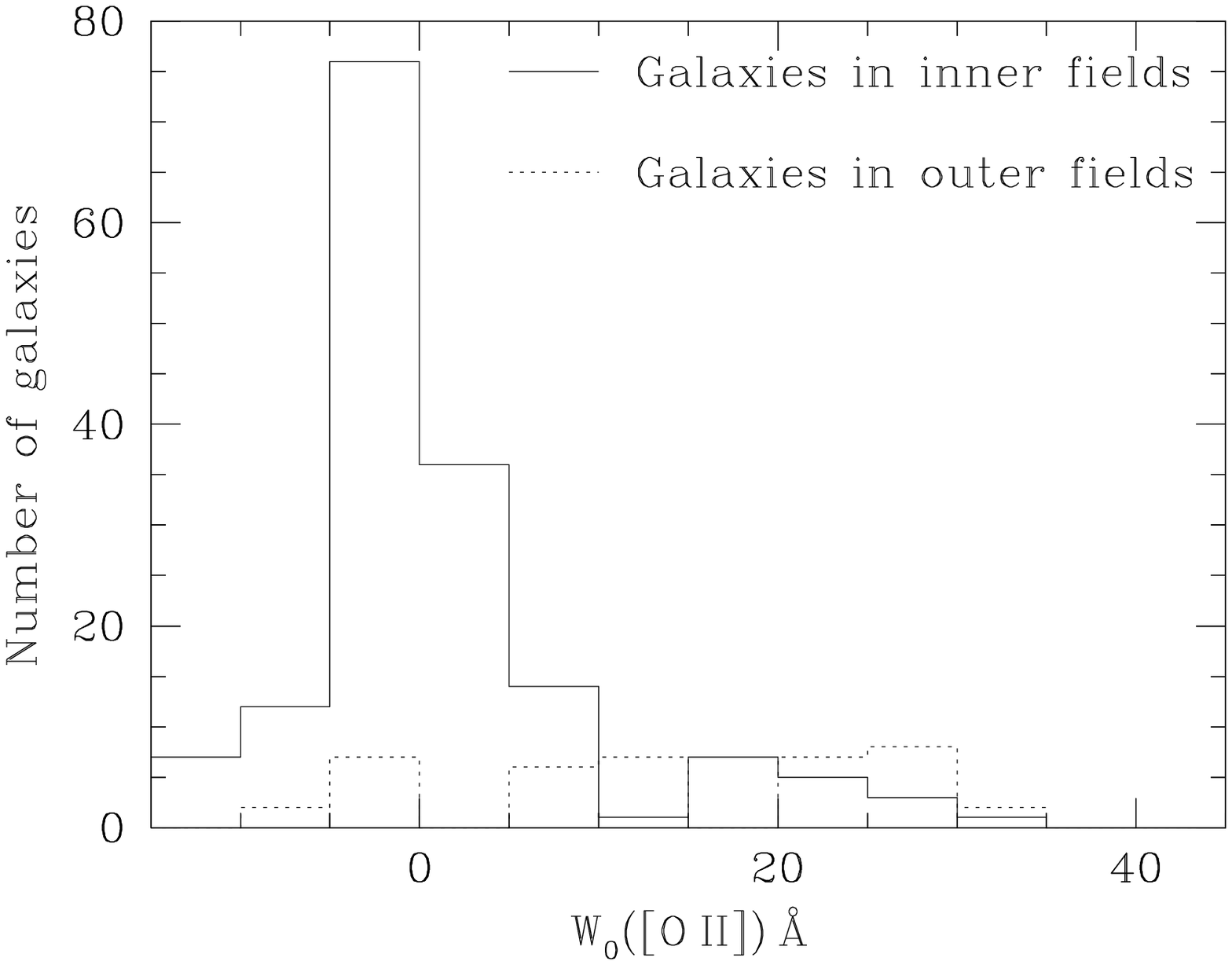}}
\caption{\small \textbf{top: }H$_\alpha$ line strength distribution for galaxies in the inner (solid line) and outer fields (dotted line). The division into inner and outer fields was performed based on the slitmasks. \textbf{bottom: }The same as in the top panel, but for [O\,II].}
\label{fig:W_0hist}
\end{figure}
The H$_\alpha$ and [O\,II] line strength distributions of galaxies in the clusters VMF73 and VMF74 are shown in Figure~\ref{fig:W_0hist}. The sample is devided into inner fields (represented by a solid line) and outer fields (dotted line), according to the slitmasks (see Sec.~\ref{sec:Spec}). This division corresponds roughly to galaxies within and beyond 1\,R$_v$, respectively. Both line strength distributions are weighted by the selection function. In the H$_\alpha$ as well as in the [O\,II] distribution, the galaxies in the inner fields show a strong peak at line strengths of $\sim0$\,\AA\ with a tail towards larger line strengths. Galaxies in the outer fields show a flat distribution in the [O\,II] line strengths and a slight peak around $\sim15$\,\AA\ in the H$_\alpha$ distribution. In the inner regions, a galaxy fraction of 19\% has H$_\alpha$ emission stronger than 5\,\AA\@, while in the outer regions this fraction is 79\%. A similar trend has been found in the [O\,II] distribution with 20\% of galaxies with W$_0$([O\,II])\,$>5$\,\AA\ in the inner fields and a fraction of 77\% of these galaxies in the outer fields.\\ 
\begin{figure}
\resizebox{\hsize}{!}{\includegraphics{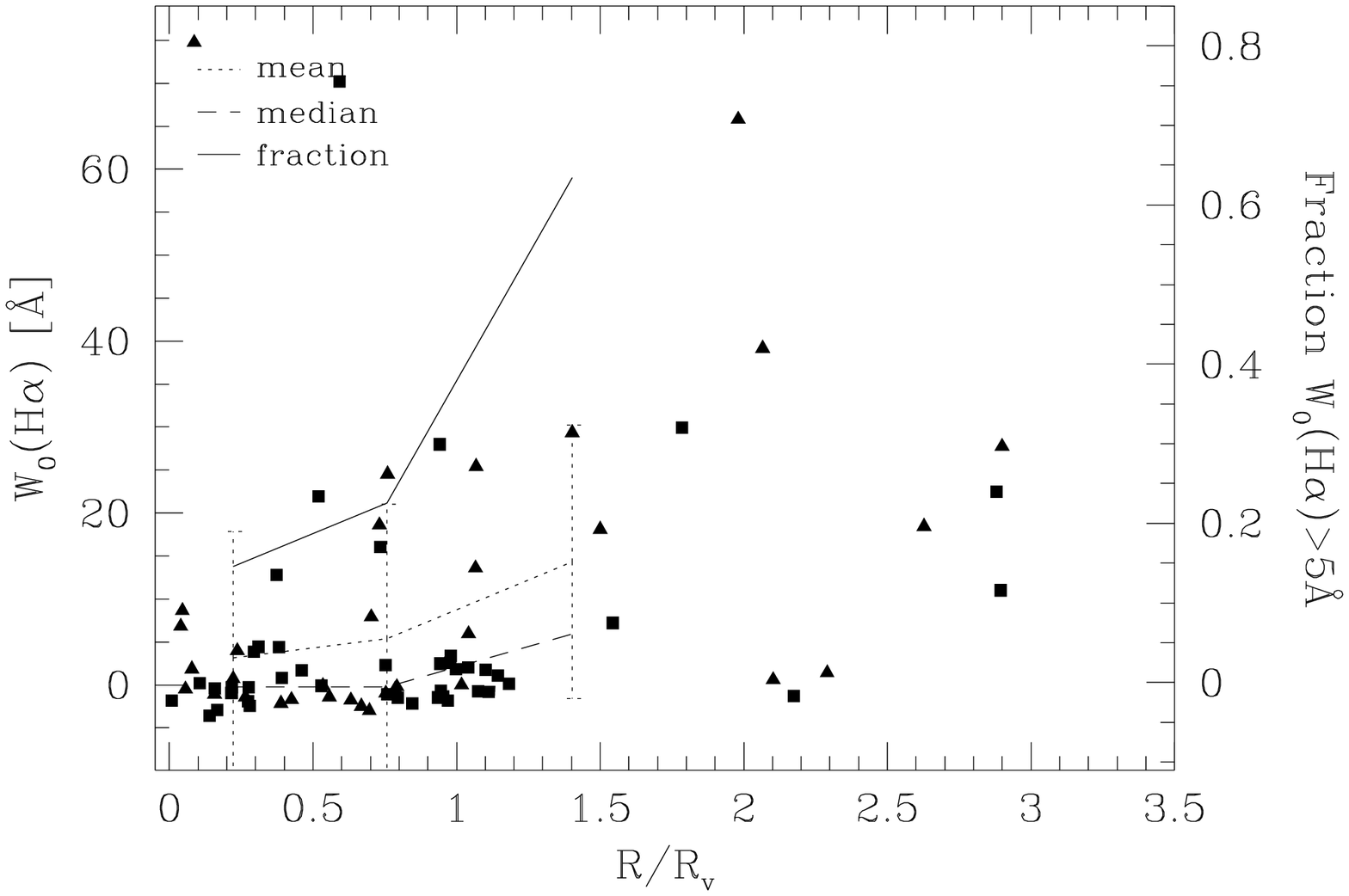}}
\resizebox{\hsize}{!}{\includegraphics{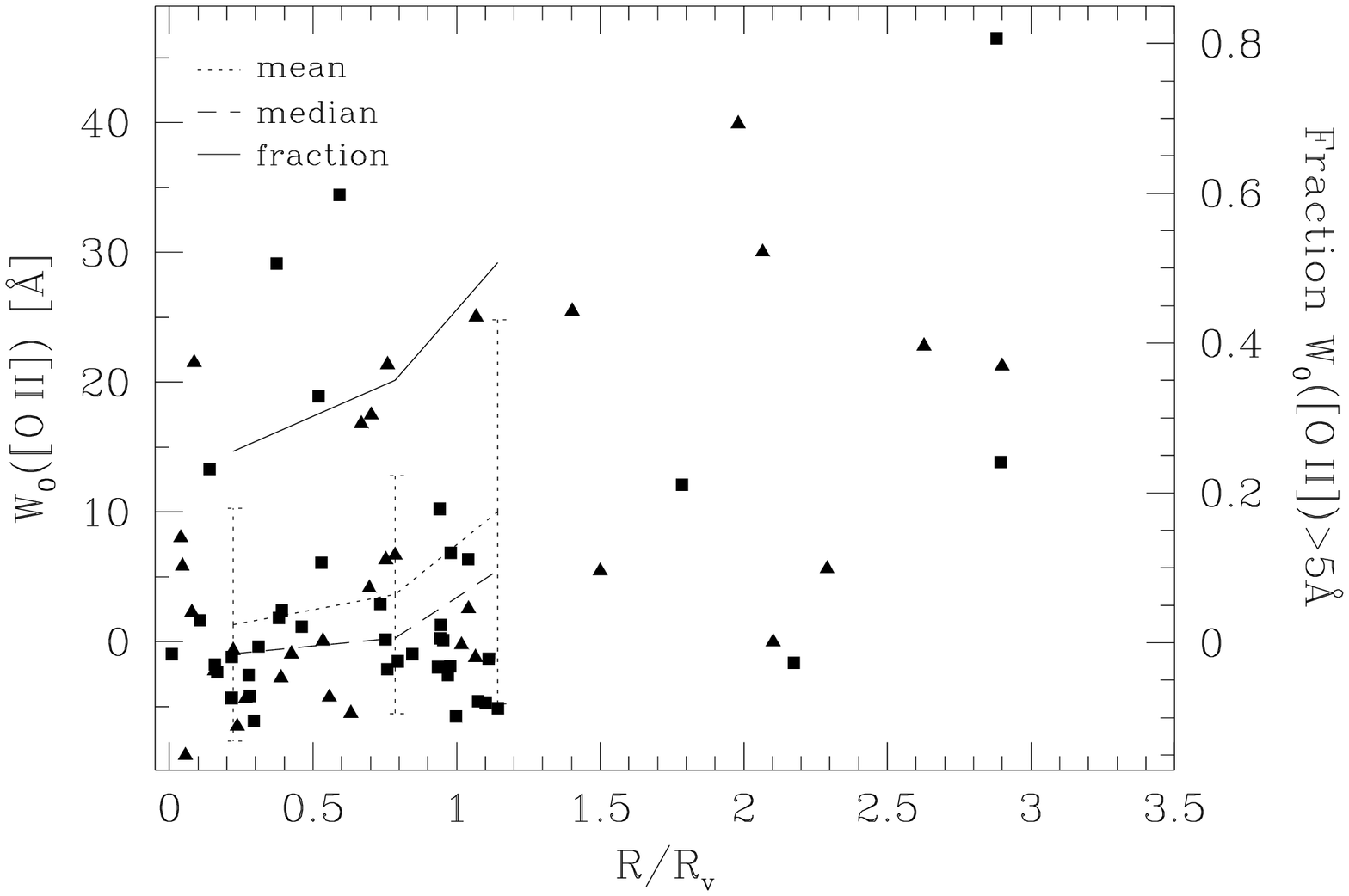}}
\caption[W$_0$ as a function of clustercentric radius]{\small \textbf{top: }Mean with 1\,$\sigma$ error bars and median of H$_{\alpha}$ line strength distribution and fraction of galaxies with W$_0$(H$_{\alpha})>5$\,\AA\ as a function of clustercentric radius. The X-ray centre is assumed to be the cluster centre. The radial bins are of varying width each containing 27 galaxies. Mean (dotted line), median (dashed line) and the fraction of galaxies with W$_0$(H$_{\alpha})>5$\,\AA\ (solid line) are weighted by the selection function. Squares and triangles represent galaxies of VMF73 and VMF74, respectively. \textbf{bottom: }The same as in the top panel, but for [O\,II].}
\label{fig:rvW0}
\end{figure}
In the top of Figure~\ref{fig:rvW0}, the mean with 1\,$\sigma$ error bars and the median of the H$_{\alpha}$ line strength distribution in the VMF73 and VMF74 sample and the fraction of galaxies with W$_0$(H$_\alpha)>5$\,\AA\ are plotted as a function of clustercentric radius with the X-ray centre assumed to be the cluster centre. The radial bins are of varying width. Each one contains 27 galaxies, the last bin comprises the last 25 galaxies and two galaxies from the second last bin. Mean (dotted line), median (dashed line) and the fraction of galaxies with an emission line strength of W$_0$(H$_\alpha)>5$\,\AA\ are weighted by the selection function. The limit of W$_0$(H$_{\alpha})=5$\,\AA\ to distinguish between passive and star forming galaxies was adopted to be consistent with \citet{GNMBG03} (see Sec.~\ref{sec:comp}). There are radial trends in the mean, median and in the fraction of star forming galaxies visible. The mean (median) of the line strength distribution increases from $\sim4\pm13$\,\AA\ ($\sim0$\,\AA\@) in the inner parts to $\sim14\pm16$\,\AA\ ($\sim6$\,\AA\@) in the outer cluster regions. Likewise, the fraction of star forming galaxies increases from $\sim15$\% in the centre to $\sim64$\% in the outer parts. 
\\
The bottom of Figure~\ref{fig:rvW0} shows the same as the top, but for [O\,II]. There are similar radial trends visible as for H$_\alpha$. The mean (median) of the line strength distribution increases from $\sim2\pm8$\,\AA\ ($\sim0$\,\AA\@) in the central parts to $\sim10\pm15$ ($\sim5$\,\AA\@) in the outer regions. In the same radial range, the fraction of star forming galaxies increases from $\sim15$\% to $\sim50$\%. 
\begin{figure}
\resizebox{\hsize}{!}{\includegraphics{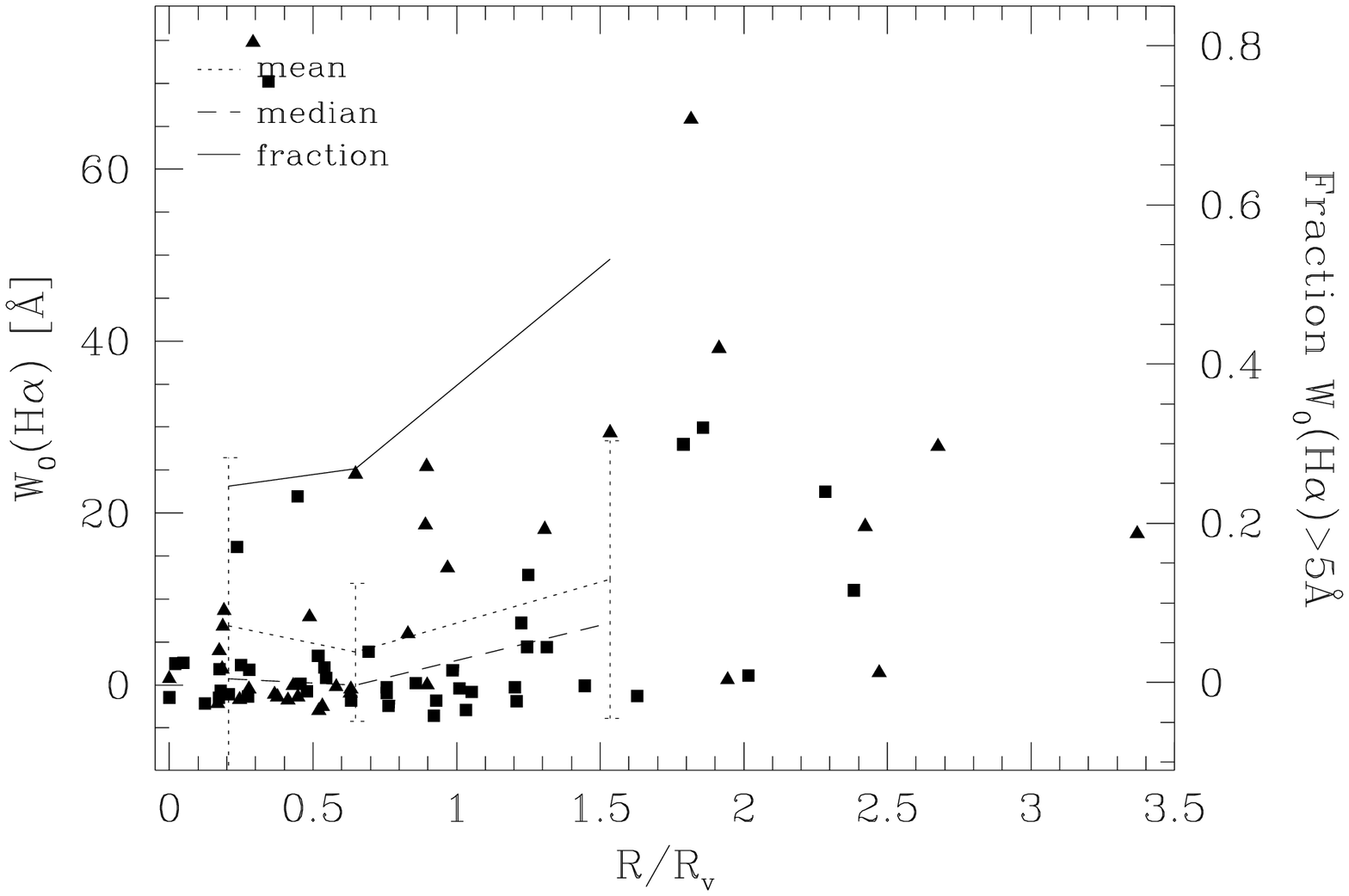}}
\resizebox{\hsize}{!}{\includegraphics{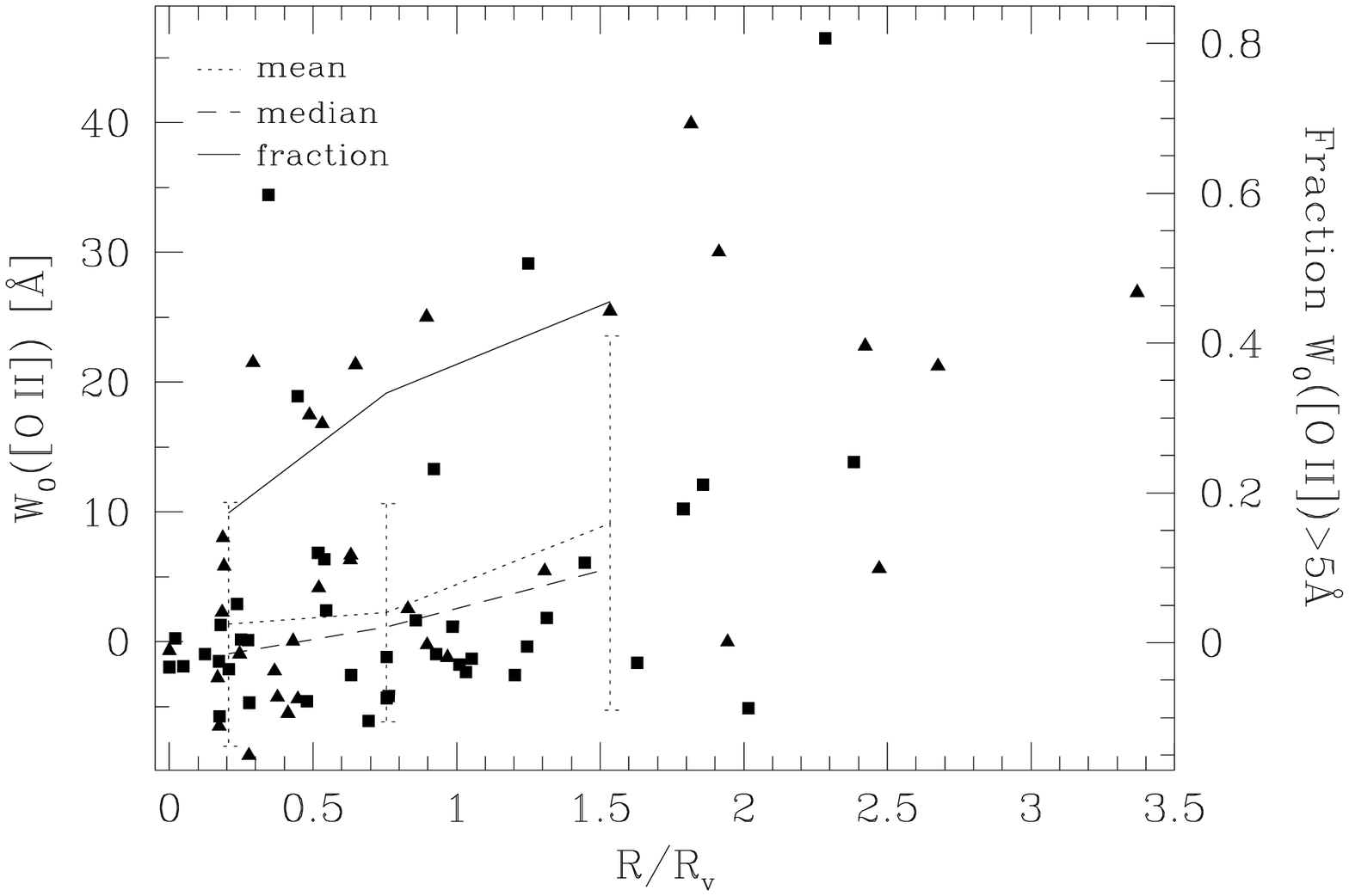}}
\caption[W$_0$ as a function of clustercentric radius]{\small The same as in Figure~\ref{fig:rvW0}, but the respective brightest cluster galaxies in the spectroscopic sample are assumed to be the cluster centre.}
\label{fig:rvIW0}
\end{figure}
Figure~\ref{fig:rvIW0} shows the same as Figure~\ref{fig:rvW0}, but with the respective $I$ band brightest cluster galaxies of the spectroscopic sample assumed to be the cluster centre. The radial dependences of the mean, median and the fraction of star forming galaxies are qualitatively similar to the ones shown in Figure~\ref{fig:rvW0}. In both the H$_\alpha$ and [O\,II] distribution, the mean, median and the star forming galaxy fraction increase with increasing clustercentric distance.

\subsubsection{Emission Line Strengths as a Function of Local Projected Galaxy Density} \label{sec:W0dens}
\begin{figure}
\resizebox{\hsize}{!}{\includegraphics{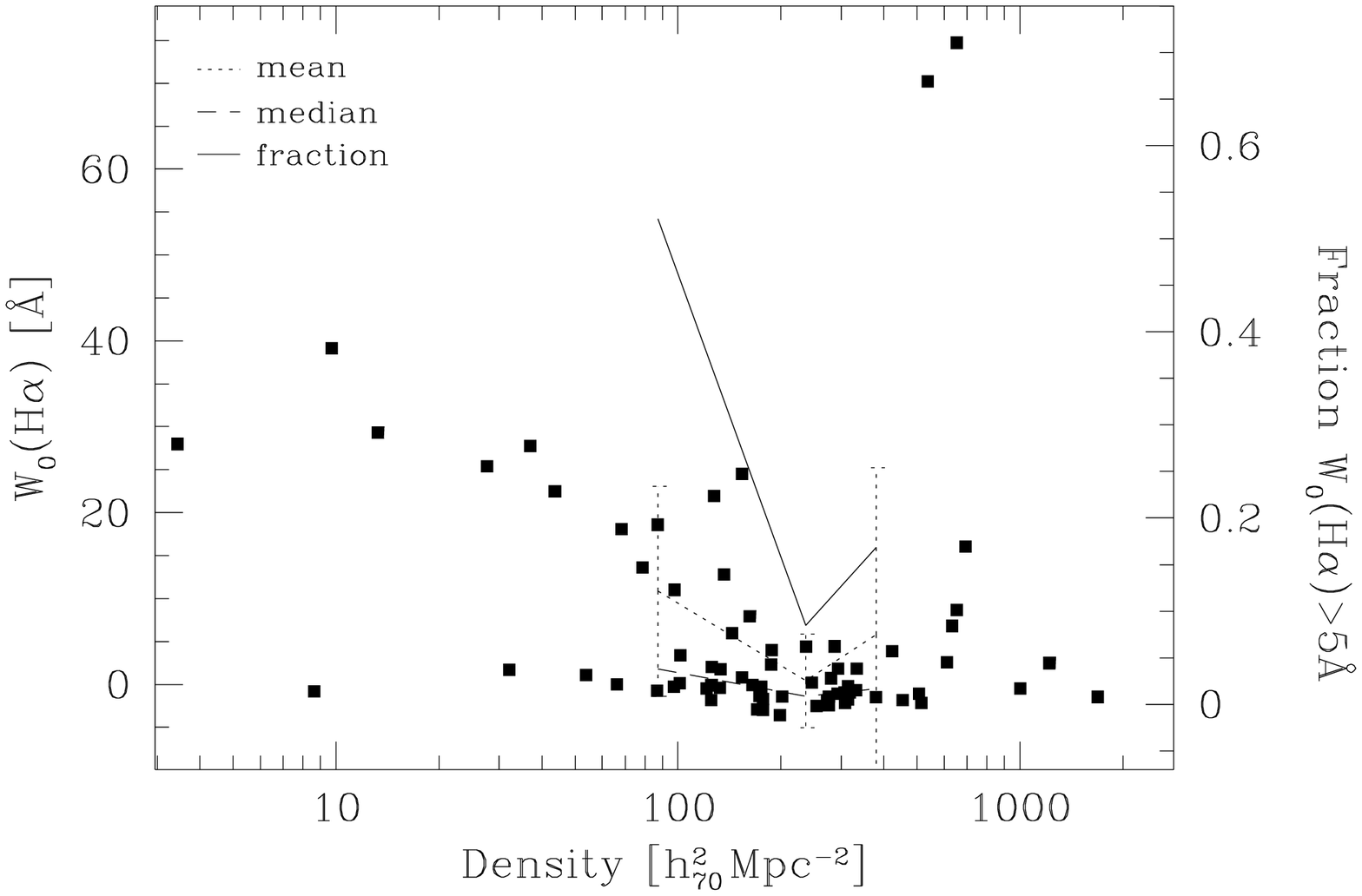}}
\resizebox{\hsize}{!}{\includegraphics{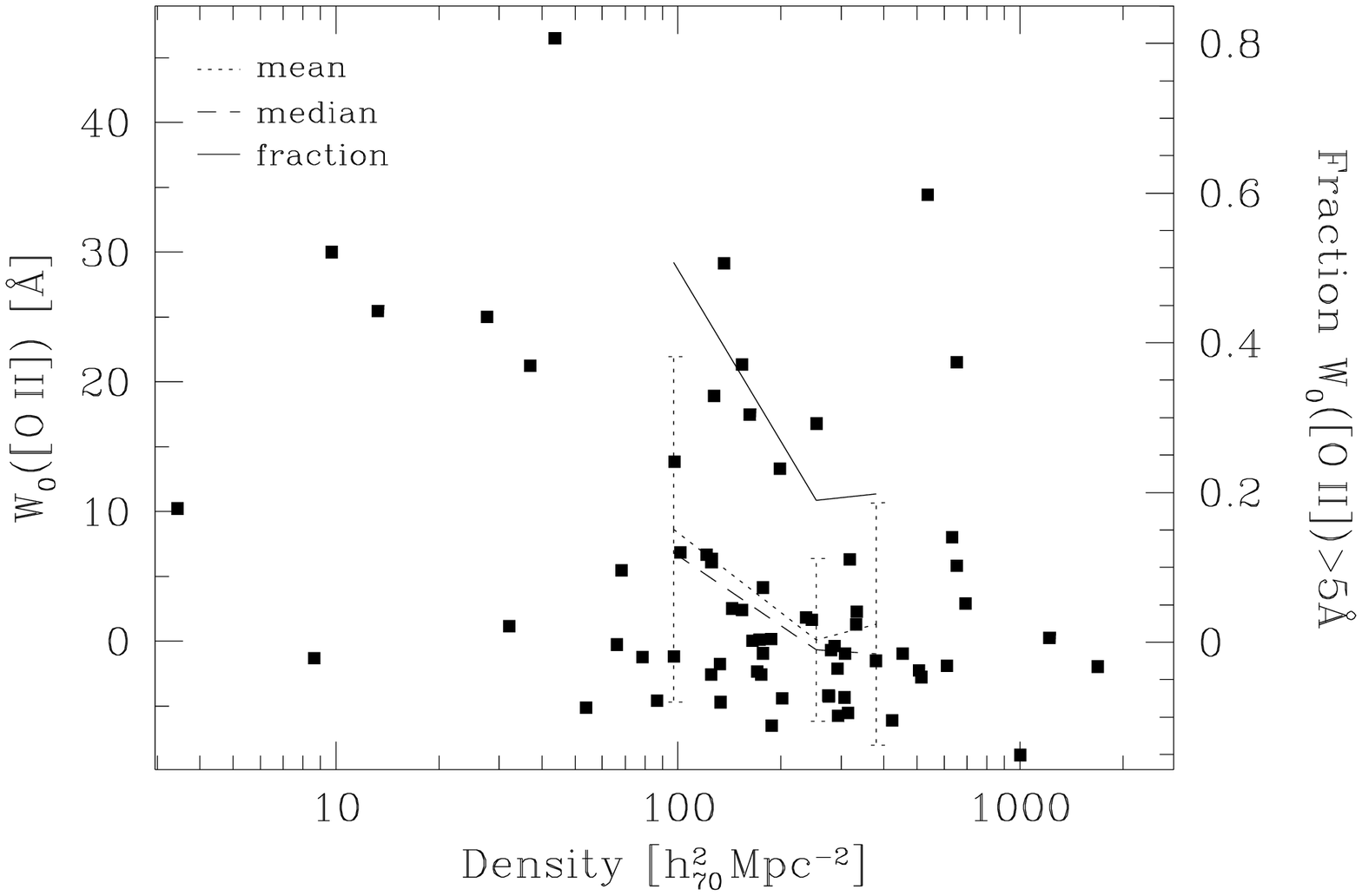}}
\caption[Line strengths of galaxies in VMF73 and VMF74 as a function of local galaxy density]{\small \textbf{Top: }Mean with 1\,$\sigma$ error bars and median of the H$_{\alpha}$ line strength distribution and fraction of galaxies with W$_0$(H$_{\alpha})>5$\,\AA\ as a function of local projected galaxy density. The radial bins are of varying width each containing 27 galaxies. Mean (dotted line), median (dashed line) and the fraction of galaxies with W$_0$(H$_{\alpha})>5$\,\AA\ (solid line) are weighted by the selection function. \textbf{Bottom: }The same as in the top panel, but for [O\,II].}
\label{fig:densEm}
\end{figure}
The line strength distributions of H$_{\alpha}$ and [O\,II] are plotted as a function of local projected galaxy density for VMF73 and VMF74 in Figure~\ref{fig:densEm}. The mean with 1\,$\sigma$ error bars and the median are represented by dotted and dashed lines, respectively. Both quantities are weighted by the selection function. The solid line shows the galaxy fractions with W$_0$(H$_{\alpha})>5$\,\AA\ and W$_0$([O\,II])\,$>5$\,\AA\ according to the scale on the right side of the figure weighted by the selection function. The density bins are of varying width and contain 27 galaxies each. The local projected galaxy density was calculated from the distance to the 5th nearest neighbour in a circular aperture of all galaxies with photometry and with a magnitude limit of $I_{tot}$\,=\,19.5 which corresponds to the magnitude limit of the spectroscopic sample. For the background correction, the number counts of \citet{LYCMS99} were adopted with a number of galaxies brighter than R$_c$\,=\,20 of 1460 per square degree. In order to account for the two clusters projected onto each other on the sky sphere, the measured local galaxy densities were divided by 2. \\
In the top of Figure~\ref{fig:densEm}, the density dependences of the H$_\alpha$ line strength distribution are shown. The mean (median) of the H$_\alpha$ distribution decreases from $\sim10\pm13$\,\AA\ ($\sim2$\,\AA\@) at lower densities to $\sim0\pm6$\,\AA\ ($\sim0$\,\AA\@) at high densities. The fraction of star forming galaxies decreases from $\sim52$\% at lower densities to $\sim9$\% in high density regimes.\\ 
The bottom of Figure~\ref{fig:densEm} shows the density trends in the [O\,II] line strength distribution. A similar trend as in the H$_\alpha$ distribution is seen. The mean (median) decreases from $\sim8\pm14$\,\AA\ ($\sim7$\,\AA\@) at low densities to $\sim0\pm7$\,\AA\ ($\sim0$\,\AA\@) at high densities. The fraction of galaxies with W$_0$([O\,II])\,$>5$\,\AA\ decreases from $\sim50$\% at lower densities to $\sim19$\% at high densities. \\
In both line strength distributions, there is an increase in the fraction of star forming galaxies in the highest density regions detected. This increase is partly but not entirely due to two galaxies with W$_0$(H$_{\alpha})>70$\,\AA\ and respective [O\,II] equivalent widths of W$_0$([O\,II])\,$=22$\,\AA\ and W$_0$([O\,II])\,$=34$\,\AA\@. These objects appear also in Figure~\ref{fig:rvW0} and Figure~\ref{fig:rvIW0} at small distances from the cluster centre. Since both the clustercentric distance and the local projected galaxy density are two-dimensional scale parameters it cannot be determined if these two galaxies reside in the centre or are projected onto it. A detailed examination of both objects did not show any signs for AGN. The colours of both objects are quite blue and are consistent with late to very late type spiral galaxies \citep{FSI95}. The bluer object has apparently a small object in its neighbourhood for which we do not have spectroscopy, so it is not clear whether it is a companion. Although we cannot draw any firm conclusions regarding a possible star forming galaxy population in the cluster centre based on only two galaxies, we cannot rule out that there is such a population. 

\subsubsection{Colour Relations}
Photometric galaxy properties are subject to changes with environment as well as spectroscopic properties. Relations between morphological type and colours of galaxies are well established and known for a long time \citep[e.g.][]{Vauco60}: Spiral galaxies with generally younger stellar populations in the disk have rather blue colours while elliptical galaxies with little or no star formation have redder colours. As galaxy morphology is related to the environment \citep[e.g.][]{Dress80,WG93}, radial gradients of colours can be expected as well.
\begin{figure}
\resizebox{\hsize}{!}{\includegraphics{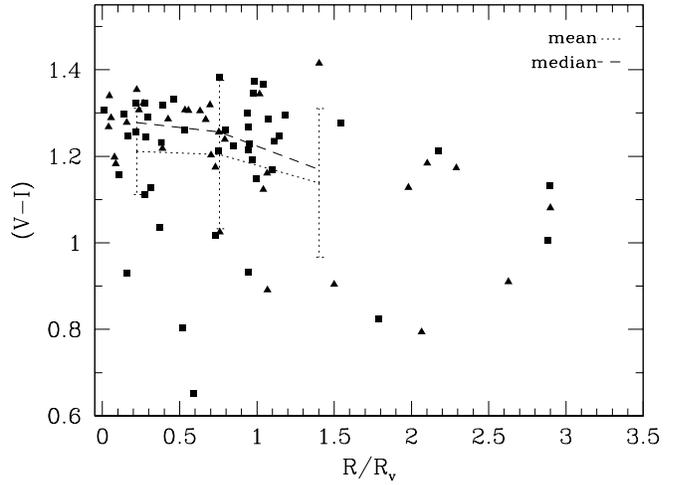}}
\caption[Galaxy colour as a function of clustercentric radius]{K-corrected observed $V-I$ colours of galaxies in VMF73 and VMF74 as a function of clustercentric radius. The respective X-ray centroids are assumed to be the cluster centres. The mean with 1\,$\sigma$ error bars and median of the colour distribution are shown as dotted and dashed lines, respectively. The mean is weighted by the selection function. Triangles and squares represent members of VMF74 and VMF73, respectively}
\label{fig:vir}
\end{figure}
In Figure~\ref{fig:vir} the k-corrected observed-frame $V-I$ colours of member galaxies of VMF73 and VMF74 are plotted against the clustercentric radius with the X-ray centre assumed to be the cluster centre. The colours are corrected by an empirical k-correction of 0.17 mag from the difference of both red sequences in the CMD. The mean with 1\,$\sigma$ error bars and the median of the colour distribution are shown as dotted and dashed lines, respectively. The mean is weighted by the selection function. Squares and triangles represent cluster members of VMF73 and VMF74, respectively. The radial bins are of varying width, each bin contains 27 galaxies. Although our data has a rather large scatter, a correlation between galaxy colour and clustercentric distance can be seen. In the central regions, the mean (median) of the colour distribution is $(V-I)\sim1.21\pm0.09$ (1.28) and becomes bluer with increasing clustercentric distance reaching final values of $(V-I)\sim1.14\pm0.17$ (1.17) in the outer regions. A colour change is visible at $\sim1$\,R$_v$ where the red sequence disappears. However, this colour break may be partly due to the small number of galaxies at large clustercentric distances.\\
\begin{figure}
\resizebox{\hsize}{!}{\includegraphics{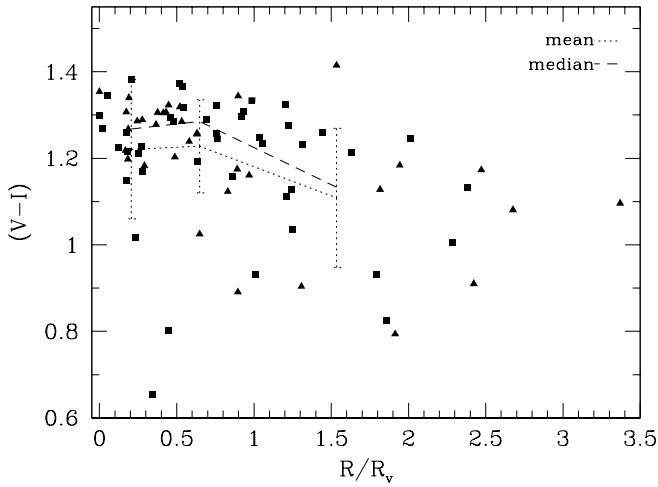}}
\caption[Galaxy colour as a function of clustercentric radius]{The same as in Figure~\ref{fig:vir}, but with the respective $I$ band brightest galaxies in the spectroscopic sample assumed to be the cluster centre.}
\label{fig:virI}
\end{figure}
Figure~\ref{fig:virI} shows the same as Figure~\ref{fig:vir}, but with the $I$ band brightest galaxies in the spectroscopic sample assumed to be the respective cluster centres. The mean and median of the colour distribution become bluer with increasing clustercentric distance, similar to the trend seen in Figure~\ref{fig:vir}. Also, a similar disappearance of the red sequence as in Figure~\ref{fig:vir} is visible.\\
\begin{figure}
\resizebox{\hsize}{!}{\includegraphics{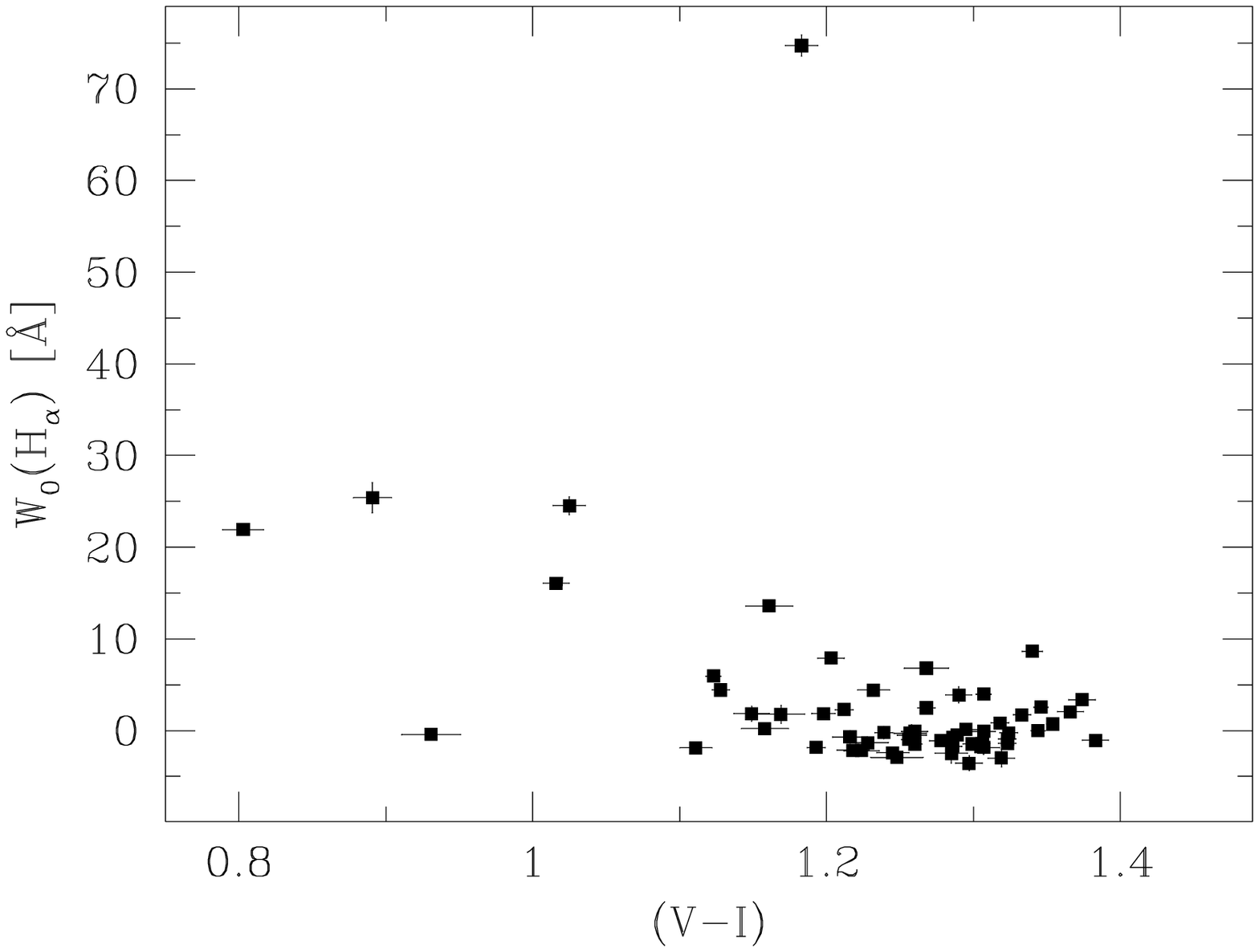}}
\resizebox{\hsize}{!}{\includegraphics{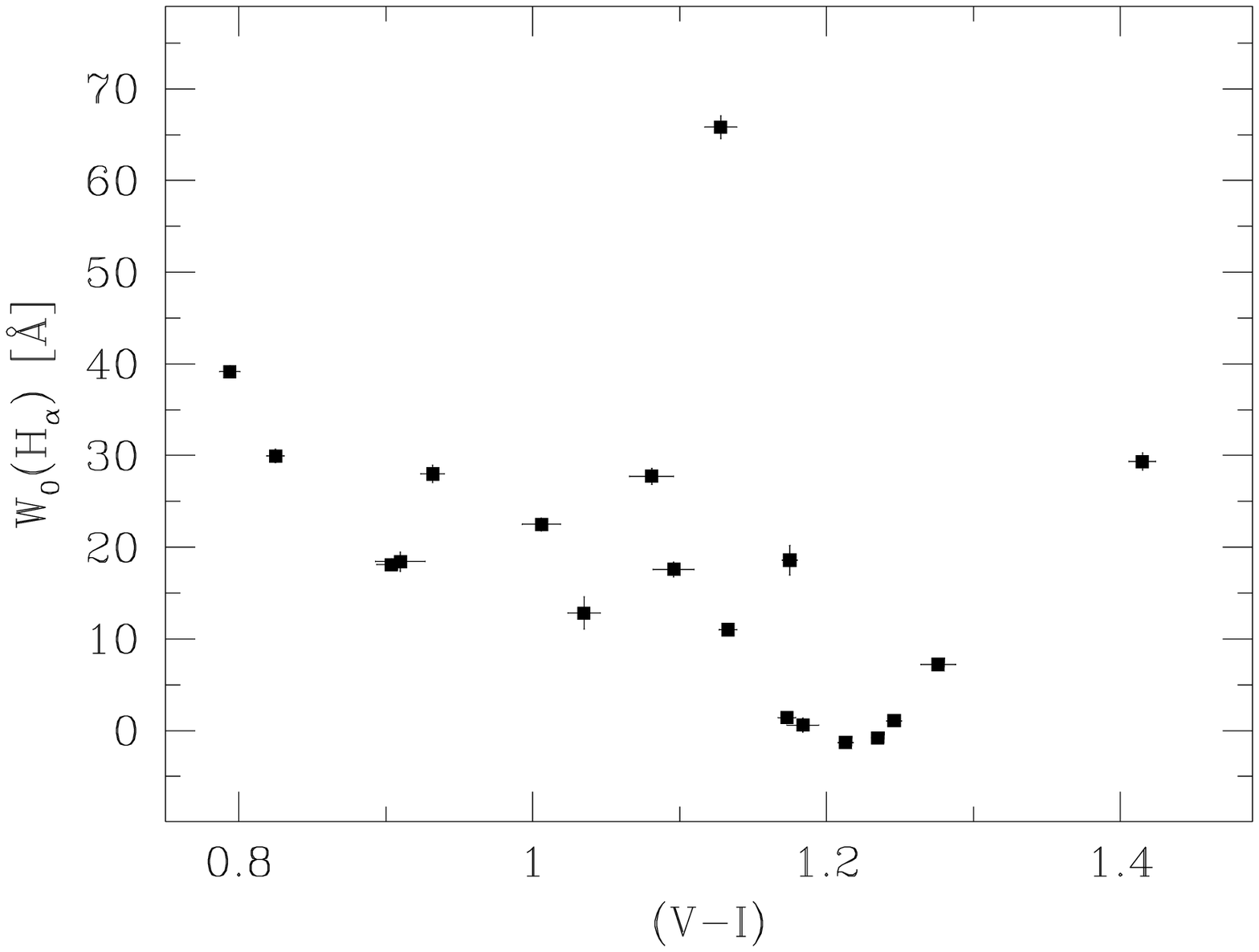}}
\caption[Line strengths as a function of colour]{\small \textbf{Top: }H$_\alpha$ line strength as a function of $V-I$ colours for the central cluster fields. \textbf{Bottom: }The same as in the top panel, but for the outer fields.}
\label{fig:vicHab}
\end{figure} 
In Figure~\ref{fig:vicHab} the relation between the H$_\alpha$ emission line strength and the k-corrected $V-I$ colour is shown. The sample is devided into an inner (top) and an outer subsample (bottom). Figure~\ref{fig:vir} and Figure~\ref{fig:rvW0} show that there are few red galaxies without emission at large clustercentric distances. In Figure~\ref{fig:vicHab}, there are some rather blue galaxies without H$_\alpha$ emission seen in the central fields. In the outer regions, there are only two galaxies with W$_0$(H$_{\alpha})<5$\,\AA\ bluer than $V-I=1.2$. This colour corresponds to late type spiral galaxies at z$=0.2$ \citep{FSI95}. There are no galaxies without H$_\alpha$ emission bluer than $V-I=1.15$ in the outer fields, whereas there are two such galaxies in the inner fields. 

\section{Discussion} \label{sec:Dis}
\subsection{Comparison with earlier work} \label{sec:comp}
The relation between star formation properties of galaxies and their environment has been investigated by various studies of the local universe \citep{LBDCB02,GNMBG03} and at intermediate redshift \citep[e.g.][]{BBSZD02}. In all of these studies, the clustercentric distance as well as the local galaxy density are used to characterize the cluster environment.\\
\citet{LBDCB02} and \citet{GNMBG03} have analysed the star formation activity of \mbox{2dFGRS} and \mbox{SDSS} galaxies in the redshift range $0.05\leq$\,z\,$\leq0.1$. Both studies derive SFRs from the H$_{\alpha}$ flux and analyse star formation properties as a function of clustercentric radius as well as of local galaxy density. \citet{GNMBG03} investigate furthermore the environmental dependence of the H$_{\alpha}$ and [O\,II] emission line strengths. The trends found in the present study are qualitatively similar to the ones detected in these earlier studies. Star formation activity decreases with decreasing clustercentric radius and increasing local galaxy density. The median of both line strength distributions in the sample of VMF73 and VMF74 increases from $\sim0$\,\AA\ in the centre to $\sim5$\,\AA\ in the outer regions. This is in good agreement with \citet{GNMBG03}. The same study finds density dependences of the star formation activity such that the median of the H$_\alpha$ and [O\,II] line strength distributions decreases with increasing local galaxy density. The density estimates used by \citet{GNMBG03} are different from the ones in the present study. \citet{GNMBG03} apply a background subtraction using redshifts while in the present analysis a statistical background subtraction was applied. As a result, the densities in Figure~\ref{fig:densEm} are much larger than those of \citet{GNMBG03} due to the difference in projection. In the VMF73 and VMF74 clusters, we see a qualitatively similar effect in H$_\alpha$ and [O\,II]; however, the sample size is too small to warrant a detailed comparison.\\  
In a redshift range comparable to the one of the present analysis, \citet{BBSZD02} have carried out a study of ten galaxy clusters with low X-ray luminosity. Galaxy clusters of low X-ray luminosity have been chosen in order to study the environmental dependence of galaxy SFRs and possible mechanisms suppressing the star formation in low mass clusters which are considered the progenitors of more massive clusters in models of hierarchical structure formation. The cluster sample of VMF73 and VMF74 shows an increase in the mean and median of the [O\,II] line strength distribution from the centre to the outer regions which is consistent with \citet{BBSZD02}. However, there is large scatter in the data of VMF73 and VMF74 and the observed trends in mean and median are of limited statistical significance. A stronger trend is found in the increase of the galaxy fraction with W$_0$([O\,II])\,$>5$\,\AA\ from $\sim15$\% in the centre to $\sim50$\% in the outer fields. \citet{BBSZD02} find an increase in the fraction of star forming galaxies from $\sim10$\% in the centre to $\sim30$\% at $\sim1$\,R$_v$. Density trends in the sample of VMF73 and VMF74 have been detected such that the mean and median of the [O\,II] line strength distribution decrease with increasing density, while \citet{BBSZD02} find that the mean reaches $\sim6$\,\AA\ only in the lowest density regions. Both studies detect similar density trends in the fraction of galaxies with W$_0$([O\,II])\,$>5$\,\AA\@.\\
As a further indicator for star formation, galaxy colours in the two clusters VMF73 and VMF74 were investigated. We find a radial trend in the colours such that with increasing radius the average colours become bluer. At large radii beyond $\sim1.5$\,R$_v$, there are few galaxies with a dominant old, red population. On the other hand, there are some rather blue galaxies without emission found in the central regions. This population may consist of galaxies which had ongoing star formation at the time of their infall into the cluster environment. On their way through the cluster, the diffuse gas in the halo may have been lost which therefore cannot replenish the fuel for star formation in the disk and the SFR slowly declines \citep{BNM00}. However, this mechanism predicts a systematically lower SFR of star forming galaxies in high density environments compared with similar galaxies in lower densities. A recent study of SDSS and 2dF data finds that the lower level of star formation in high density environments is largely due to the smaller fraction of star forming galaxies in these environments, while there is little difference in the star forming galaxy populations in high and low density regimes \citep{BEMLB04}. The sample of VMF73 and VMF74 is too small to draw any strong conclusions from the population of blue galaxies without H$_\alpha$ emission. Figure~\ref{fig:vicHab} suggests that there is a small difference in the H$_\alpha$ emission between blue, star forming galaxies in the inner and outer fields. Furthermore, in the VMF73 and VMF74 sample a distinctive colour change is visible at a clustercentric radius of $\sim1$\,R$_v$ where the red sequence of both clusters disappears. This feature is reminiscent of the work of \citet{KSNOB01} who find an abrupt colour change at a characteristic density of 2 galaxies h$^2_{50}$ Mpc$^{-2}$ in the cluster Cl~0939+4713 at z$=0.41$.

\subsection{Physical Implications}
The aim of this investigation has been to explore the star formation properties of galaxies over a wide range of densities with a special focus on galaxies in large distances from the cluster core. The difference between the present study and the two previous ones cited above is that we investigated cluster galaxies in a higher redshift regime than did \citet{GNMBG03} analysing \mbox{SDSS} data. Furthermore, this work reaches out to larger clustercentric distances than the study of X-ray faint clusters at intermediate redshift \citep{BBSZD02}. In regions as far out as $\sim3-4$\,R$_v$ from the cluster centre, galaxies from the surrounding field are thought to fall continuously into the cluster attracted by its gravitational potential and experience for the first time the influence of the cluster environment. Therefore, galaxies in these infall regions which denote a transition between the low density field and the cluster are expected to have intermediate properties between the field and cluster galaxy population. The results of \citet{GNMBG03} and \citet{BBSZD02} are consistent with this picture in finding an increase in line strengths and SFRs with increasing clustercentric distance and decreasing density. The present study finds similar radial and density dependences of galaxy star formation. The change of galaxy colours with radius is a further indication for some kind of transformation of cluster galaxies. The observed disappearance of the red sequence at a radius of $\sim1$\,R$_v$ is similar to an abrupt colour break reported by \citet{KSNOB01} at a characteristic density of 2 galaxies h$^2_{50}$ Mpc$^{-2}$. \citet{KSNOB01} find that this density corresponds to subclumps outside the cluster centre. This may be a further hint that galaxy transformation first occurs in groups and subclumps far from the centre.\\ 
Several processes are being discussed in the literature as candidates for suppressing star formation in cluster galaxies. The radial dependence of star formation properties suggests that this mechanism starts to act about 2.5\,R$_v$ from the cluster core and possibly even beyond. This result rules out mechanisms that are effective only in the highest density regimes, such as ram pressure stripping, as being solely responsible for the decreasing SFRs throughout the cluster. Although more than one process may be involved and ram pressure stripping may still play a role in the cluster core, it is likely that in the outer regions a more subtle mechanism is at work which does not halt the star formation abruptly. Apparently, this mechanism starts to affect galaxies relatively early after their first contact with the cluster environment, for even galaxies far out of the central regions show a reduced star formation. The scenario of strangulation \citep{BNM00} in which the diffuse gas in the dark halo of an infalling galaxy is lost may be an explanation for the reduced SFRs in cluster galaxies. As discussed above, this mechanism predicts lower SFRs of blue galaxies in high density regions compared with blue galaxies in lower density regimes. However, based on the blue galaxy population in the VMF73 and VMF74 cluster sample alone, we cannot draw any strong conclusions about the mechanism suppressing star formation in the cluster environment. The observed decrease in SFRs is compatible with strangulation, although it is possible that other effects such as galaxy-galaxy interactions play a role or were important in the past and influence galaxy evolution in small groups even before a galaxy enters the cluster environment.

\subsection{Outlook}
The reduction and analysis of the remaining four galaxy clusters will increase the sample size and enable us to investigate the environmental dependence of galaxy star formation properties with more precision and statistical significance. With the complete sample of six galaxy clusters, both the SFR-radius relation and the SFR-density relation can be studied without the limitations of low number statistics and thus help us to understand the evolution of galaxies in the cluster environment.\\
Additionally, it would be interesting to study the dependence of star formation activity on the X-ray luminosity of clusters. Groups and low mass X-ray faint clusters are considered the progenitors of more massive clusters in hierarchical merging models. Studying clusters over a range of masses is therefore an important step to investigate the growth of structure in the universe.

\section{Summary}
We have observed three sky fields selected from the XDCS \citep{GBCZ04}, each containing two galaxy clusters, with multiobject spectroscopy. In the present paper, the two clusters contained in the field R285 were analysed. We investigated the star formation activity of cluster galaxies based on the line strengths of H$_\alpha$ and [O\,II] with special emphasis on galaxies far from the cluster centre. The results of the analysis can be summarized as follows:
\begin{enumerate}
\item
The two clusters VMF73 and VMF74 in field R285 lie at redshifts of z\,$=0.18$ and z\,$=0.25$ with velocity dispersions of 442\,km\,s$^{-1}$ and 671\,km\,s$^{-1}$, respectively. The velocity dispersions of both clusters are consistent with Gaussians and are in good agreement with the local L$_X$--$\sigma$ relation found in samples of \citet{DSJFV93,Marke98} and \citet{XW00}. Within 3\,$\sigma$, the number of spectroscopically observed member galaxies is 44 in VMF73 and 35 in VMF74. The approximate virial radii of VMF73 and VMF74 are 1.67\,Mpc and 1.21\,Mpc, respectively, in our cosmology.
\item
The fraction of emission line galaxies with W$_0$([O\,II])\,$>5$\,\AA\ in the cluster sample is 31\% in total, weighted by the selection function. Dividing the cluster sample into inner and outer masks, the weighted fraction of galaxies with W$_0$([O\,II])\,$>5$\,\AA\ is 20\% within $\sim1$\,R$_v$ and 77\% in the outer fields. The weighted fraction with W$_0$(H$_\alpha)>5$\,\AA\ is 19\% in the inner fields and 79\% in the outer regions. 
\item
The line strength distributions of H$_\alpha$ and [O\,II] in the VMF73 and VMF74 sample show radial dependences in such a way that the mean and median and the star forming galaxy fraction increase with clustercentric radius. In the H$_\alpha$ distribution, the mean (median) increases from $\sim4\pm13$\,\AA\ ($\sim0$\,\AA\@) in the inner parts to $\sim14\pm16$\,\AA\ ($\sim6$\,\AA\@) in the outer cluster regions. The fraction of galaxies with W$_0$(H$_\alpha)>5$\,\AA\ increases from $\sim15$\% in the centre to $\sim64$\% in the outer parts. The mean (median) of the [O\,II] line strength distribution increases from $\sim2\pm8$\,\AA\ ($\sim0$\,\AA\@) in the central parts to $\sim10\pm15$ ($\sim5$\,\AA\@) in the outer regions. In the same radial range, the fraction of galaxies with W$_0$([O\,II])\,$>5$\,\AA\ increases from $\sim15$\% to $\sim50$\%.
\item
Density dependences of the H$_{\alpha}$ and [O\,II] emission line strengths in VMF73 and VMF74 have been detected such that the mean, median and the star forming galaxy fraction decrease with increasing local projected galaxy density. The mean (median) of the H$_\alpha$ distribution decreases from $\sim10\pm13$\,\AA\ ($\sim2$\,\AA\@) at lower densities to $\sim0\pm6$\,\AA\ ($\sim0$\,\AA\@) at high densities. The fraction of galaxies with W$_0$(H$_\alpha)>5$\,\AA\ decreases from $\sim52$\% at lower densities to $\sim9$\% in high density regimes. In the [O\,II] distribution, the mean (median) decreases from $\sim8\pm14$\,\AA\ ($\sim7$\,\AA\@) at low densities to $\sim0\pm7$\,\AA\ ($\sim0$\,\AA\@) at high densities. The fraction of galaxies with W$_0$([O\,II])\,$>5$\,\AA\ decreases from $\sim50$\% at lower densities to $\sim19$\% at high densities.
\item
A radial trend in the observed-frame $V-I$ colours has been detected. The mean and median of the galaxy colour distribution in the VMF73 and VMF74 sample decrease from $(V-I)\sim1.30\pm0.08$ at 0.2\,R$_v$ to bluer colours and reach $(V-I)\sim1.04\pm0.12$ (mean) and $(V-I)\sim1.15$ (median) at 2.6\,R$_v$. At a clustercentric distance of $\sim1$\,R$_v$, the red sequence is observed to disappear. 
\end{enumerate}  

\begin{acknowledgements}
We would like to thank A. Aguirre and the Calar Alto local staff for efficient support during the observations. We thank Prof. K. J. Fricke for support and A. B\"ohm for helpful discussions. This work has been supported by the Volkswagen Foundation (I/76\,520) and the Deutsche Forschungsgemeinschaft (ZI\,663/5--1).
\end{acknowledgements}

\bibliographystyle{aa}
\bibliography{/data/burray/bettinag/bib/abb,/data/burray/bettinag/bib/all}

\appendix
\section{Data Table} 
\begin{table*}
\centering
\caption{\small Object ID, J2000 coordinates, redshift, equivalent widths of [O\,II] and H$_\alpha$ with errors and corresponding cluster for all member galaxies with spectroscopy. The spectrum r2851\_14 is corrupted in the blue which made an [O\,II] measurement in this galaxy impossible.}
\label{tab:spectab}
\begin{tabular}{c c c c c c c c c}
\hline
object ID & R.A.&DEC &z&W$_0$([O\,II])& $\Delta$W$_0$([O\,II]) & W$_0$(H$_\alpha$) &$\Delta$W$_0$(H$_\alpha$) & cluster\\
\hline
 r2811\_20& 9:44: 3.215& 16:39:48.59&0.2557&  -4.7067&   3.0750&   1.7857&   0.9953&VMF73 \\
 r2811\_24& 9:44: 1.379& 16:38: 1.11&0.2536&  -4.5978&   1.3008&  -0.7676&   0.3461&VMF73\\
 r2811\_25& 9:43:59.685& 16:37:30.18&0.2541&   6.3543&   1.5242&   2.0627&   0.4611&VMF73 \\
 r2811\_23& 9:43:59.527& 16:38:29.82&0.1783&   2.5207&   0.9593&   5.9572&   0.2739&VMF74\\
 r2811\_16& 9:43:58.380& 16:41: 9.65&0.2527&   0.2592&   0.6512&   2.4747&   0.2478&VMF73\\
 r2811\_19& 9:43:58.818& 16:40: 2.38&0.2538&   1.2676&   2.6258&  -0.6907&   0.6743&VMF73\\
 r2811\_14& 9:43:58.757& 16:42: 2.56&0.1824&  -5.5232&   1.7171&  -1.7689&   0.3677&VMF74\\
 r2811\_10& 9:43:55.556& 16:43:34.68&0.1788&  -0.9383&   1.3420&  -1.7080&   0.3509&VMF74\\
 r2811\_18& 9:43:53.520& 16:40:23.14&0.2529&  -2.1270&   1.1838&  -1.0665&   0.3490&VMF73\\
 r2811\_13& 9:43:53.021& 16:42:48.20&0.1784&  -2.7790&   2.0136&  -2.1479&   0.5918&VMF74\\
 r2811\_06& 9:43:52.614& 16:44:40.11&0.2538&   6.8425&   1.1742&   3.3899&   0.3720&VMF73\\
 r2811\_11& 9:43:49.126& 16:43:21.27&0.1809&  -0.6746&   0.6510&   0.7147&   0.1845&VMF74\\
 r2811\_07& 9:43:46.729& 16:44:25.27&0.1795&   2.2640&   1.2236&   1.8584&   0.4123&VMF74\\
 r2811\_08& 9:43:45.157& 16:44: 5.69&0.1788&   8.0074&   1.8511&   6.8131&   0.5142&VMF74\\
 r2811\_05& 9:43:44.494& 16:44:54.27&0.1801&  21.5007&   1.1904&  74.7323&   1.1308&VMF74\\
 r2811\_01& 9:43:44.470& 16:46: 5.36&0.1784&  -4.4120&   0.7369&  -1.3891&   0.2486&VMF74\\
 r2811\_22& 9:43:43.315& 16:39:18.56&0.1770&  21.3541&   2.3289&  24.5442&   0.9740&VMF74\\
 r2811\_03& 9:43:43.514& 16:45:20.16&0.1803&  -2.2580&   0.9715&  -1.0987&   0.3454&VMF74\\
 r2812\_05& 9:44: 5.008& 16:38:34.35&0.2579& -10.6131&   3.4720&   0.1310&   0.4503&VMF73\\
 r2812\_16& 9:44: 1.022& 16:42: 4.17&0.1775&  17.4814&   4.2423&   7.9112&   0.5457&VMF74\\
 r2812\_09& 9:44: 0.321& 16:40:11.54&0.2487&  -5.7794&   3.7343&   1.8457&   0.8243&VMF73\\
 r2812\_12& 9:43:59.374& 16:41: 9.92&0.2570&  -1.8901&   1.4455&   2.5585&   0.2619&VMF73\\
 r2812\_14& 9:43:53.579& 16:41:43.22&0.2529&  -1.5178&   1.2966&  -1.4753&   0.2946&VMF73\\
 r2812\_17& 9:43:45.135& 16:42:46.34&0.1813&  -6.5287&   1.6785&   3.9807&   0.2401&VMF74\\
 r2812\_22& 9:43:43.459& 16:44:31.85&0.1810&  -8.7844&   3.0857&  -0.4729&   0.3618&VMF74\\
 r2812\_21& 9:43:44.867& 16:44: 2.25&0.1793&   5.8361&   1.1055&   8.6601&   0.2533&VMF74\\
 r2812\_02& 9:43:43.064& 16:37:36.01&0.1813&  -0.2660&   1.3675&   0.0032&   0.2233&VMF74\\
 r2821\_02& 9:43:58.089& 16:41:17.05&0.2530&  -1.9685&   1.4632&  -1.4519&   0.3745&VMF73\\
 r2821\_03& 9:43:56.389& 16:39:57.57&0.1799& -11.6547&   3.9628&  -0.2161&   0.6952&VMF74\\
 r2821\_06& 9:43:51.143& 16:37:26.30&0.1749&  25.0298&   2.8753&  25.4038&   1.6262&VMF74\\
 r2821\_07& 9:43:50.041& 16:39:54.78&0.1796&   4.1322&   4.0803&  -2.9761&   0.9504&VMF74\\
 r2821\_08& 9:43:48.710& 16:40:39.14&0.2540&  34.4289&   0.8362&  70.2181&   1.5790&VMF73\\
 r2821\_09& 9:43:47.698& 16:39:10.28&0.1797&   6.6599&   6.5108&  -0.4738&   1.1184&VMF74\\
 r2821\_10& 9:43:46.040& 16:39:54.53&0.1776&  16.7840&   7.6416&  -2.4824&   1.0909&VMF74\\
 r2821\_11& 9:43:44.532& 16:39:19.87&0.1766&   6.3128&   2.1436&  -0.9577&   0.4898&VMF74\\
 r2821\_12& 9:43:43.004& 16:40:34.58&0.2573&   2.4168&   1.4702&   0.8388&   0.4232&VMF73\\
 r2821\_17& 9:43:36.349& 16:36:57.38&0.2588&   1.1436&   2.1093&   1.7137&   0.5961&VMF73\\
 r2821\_14& 9:43:40.074& 16:39:23.66&0.2586&  -6.1052&   4.1495&   3.8824&   0.8789&VMF73\\
 r2821\_20& 9:43:33.631& 16:39: 6.81&0.2530&  13.3054&   2.7992&  -3.5572&   0.8234&VMF73\\
 r2821\_21& 9:43:32.427& 16:40: 1.03&0.2539&  -0.9675&   2.2326&  -1.8471&   0.7725&VMF73\\
 r2821\_27& 9:43:23.531& 16:39:46.40&0.2567&  -0.3700&   1.3343&   4.4400&   0.5386&VMF73\\
 r2821\_29& 9:43:19.340& 16:38: 8.64&0.2573&   6.1021&   1.6261&  -0.0986&   0.6769&VMF73\\
\end{tabular}
\end{table*}
\begin{table*}
\centering
\begin{tabular}{c c c c c c c c c}
\hline
object ID & R.A.&DEC &z&W$_0$([O\,II])& $\Delta$W$_0$([O\,II]) & W$_0$(H$_\alpha$) &$\Delta$W$_0$(H$_\alpha$) & cluster\\
\hline
 r2822\_01& 9:43:58.931& 16:39:22.06&0.2561&   0.1120&   0.9628&  -1.3375&   0.3376&VMF73\\
 r2822\_03& 9:43:56.353& 16:36:51.10&0.2552&  -2.5866&   0.6531&  -1.8366&   0.3713&VMF73\\
 r2822\_04& 9:43:55.897& 16:40:36.04&0.2551&  -0.9627&   1.0819&  -2.1862&   0.3375&VMF73\\
 r2822\_05& 9:43:53.380& 16:39:59.19&0.2511&   0.1605&   0.4556&   2.2797&   0.1815&VMF73\\
 r2822\_06& 9:43:51.720& 16:41:45.08&0.2529&   2.8820&   0.5106&  16.0670&   0.3112&VMF73\\
 r2822\_07& 9:43:49.727& 16:40:51.40&0.1805&  -4.2757&   0.5787&  -1.3949&   0.1526&VMF74\\
 r2822\_09& 9:43:45.555& 16:41:30.97&0.2517&  18.8964&   0.6678&  21.9371&   0.5421&VMF73\\
 r2822\_10& 9:43:44.153& 16:40:47.14&0.1795&   0.0361&   0.8027&  -0.0852&   0.2079&VMF74\\
 r2822\_14& 9:43:38.759& 16:38:55.57&0.2533&  -4.2059&   0.9098&  -2.4142&   0.2919&VMF73\\
 r2822\_15& 9:43:37.970& 16:39:32.66&0.2574&  -4.3497&   0.5612&  -0.9239&   0.1728&VMF73\\
 r2822\_16& 9:43:36.803& 16:41: 2.71&0.2552&  -1.1774&   0.8351&  -0.2862&   0.2456&VMF73\\
 r2822\_17& 9:43:34.091& 16:40:36.17&0.2500&   1.6263&   0.8720&   0.2113&   0.3482&VMF73\\
 r2822\_13& 9:43:39.732& 16:37:22.52&0.1795&  -1.2232&   1.8434&  13.6049&   0.4373&VMF74\\
 r2822\_19& 9:43:30.578& 16:38:56.01&0.2529&  -2.3637&   1.2785&  -2.9096&   0.4978&VMF73\\
 r2822\_20& 9:43:29.646& 16:40:56.82&0.2578&  -1.7712&   0.6528&  -0.4303&   0.4010&VMF73\\
 r2822\_22& 9:43:25.342& 16:39: 7.25&0.2549&  -2.5926&   0.5217&  -0.2400&   0.1746&VMF73\\
 r2822\_23& 9:43:24.510& 16:39:52.15&0.2614& -13.1422&   1.2640&  -1.8871&   0.5062&VMF73\\
 r2822\_25& 9:43:22.064& 16:39: 7.95&0.2504&   1.8265&   0.6719&   4.4135&   0.2738&VMF73\\
 r2831\_01& 9:43:25.504& 16:43:37.95&0.1785& -15.3682&   7.4113&  18.5928&   1.5945&VMF74\\
 r2831\_03& 9:43:22.885& 16:41:14.85&0.2475&  29.1424&   5.4647&  12.8329&   1.7592&VMF73\\
 r2831\_09& 9:43: 8.925& 16:41:44.84&0.1799&  25.4745&   3.0314&  29.3486&   0.9570&VMF74\\
 r2831\_10& 9:43: 8.025& 16:42:45.40&0.2554&  10.2344&   1.0323&  28.0087&   0.9258&VMF73\\
 r2831\_13& 9:43: 1.490& 16:42:27.79&0.2560&  -5.1153&   1.2809&   1.0695&   0.3672&VMF73\\
 r2831\_20& 9:42:44.142& 16:45:34.92&0.1795&   5.6379&   1.1953&   1.4298&   0.5835&VMF74\\
 r2841\_17& 9:44:23.762& 16:31:47.11&0.2504&  -1.6194&   0.9967&  -1.2897&   0.2258&VMF73\\
 r2841\_08& 9:44:40.098& 16:30:59.76&0.1848&  21.2407&   1.9546&  27.7546&   0.8711&VMF74\\
 r2841\_07& 9:44:41.052& 16:29:19.52&0.2500&  46.5020&   2.5282&  22.5031&   0.7267&VMF73\\
 r2841\_20& 9:44:16.292& 16:28:47.48&0.1798&  22.7927&   2.8351&  18.4427&   1.0485&VMF74\\
 r2841\_03& 9:44:54.152& 16:28: 0.61&0.1833&  26.9221&   8.7519&  17.5820&   0.8333&VMF74\\
 r2841\_10& 9:44:36.525& 16:27:31.55&0.2538&  13.8636&   0.7475&  11.0069&   0.2843&VMF73\\
 r2851\_14& 9:44: 4.693& 16:32:49.30&0.2517&    --&  --&   7.2167&   0.6526&VMF73\\
 r2851\_19& 9:43:59.003& 16:35: 2.83&0.1810&   5.4581&   1.6128&  18.1029&   0.6305&VMF74\\
 r2851\_11& 9:43:52.550& 16:31:20.41&0.1818&  39.9267&   1.3898&  65.8269&   1.2526&VMF74\\
 r2851\_17& 9:43:52.248& 16:34: 0.81&0.2520&  -1.3208&   0.5579&  -0.7933&   0.1892&VMF73\\
 r2851\_04& 9:43:50.600& 16:28:20.55&0.2501&  12.0806&   0.5646&  29.9664&   0.7405&VMF73\\
 r2851\_09& 9:43:50.527& 16:30:28.09&0.1812&  -0.0363&   2.1135&   0.6131&   0.7490&VMF74\\
 r2851\_10& 9:43:48.575& 16:30:39.77&0.1796&  30.0205&   0.6615&  39.1525&   0.6419&VMF74\\
\end{tabular}
\end{table*}

\end{document}